\documentclass[fleqn,usenatbib]{mnras}

\usepackage{newtxtext,newtxmath}

\usepackage[T1]{fontenc}
\usepackage{ae,aecompl}
\usepackage{pdflscape}

\usepackage{longtable} 
\usepackage{pdflscape} 

\usepackage{graphicx}	
\usepackage{amsmath}	

\newcommand{\thisstar}{K2-111}
\newcommand{\planetb}{K2-111\,b}
\newcommand{\planetc}{K2-111\,c}
\newcommand{\msun}{M$_\odot$}
\newcommand{\rsun}{R$_\odot$}
\newcommand{\mearth}{M$_\oplus$}
\newcommand{\rearth}{R$_\oplus$}

\title[K2-111 system]{K2-111: an old system with two planets in near-resonance\thanks{Based on Guaranteed Time Observations collected at the European Southern Observatory under ESO programmes 1102.C-0744 and 1104.C-0350 by the ESPRESSO Consortium and at the Italian Telescopio Nazionale Galileo (TNG) by the HARPS-N Collaboration.}}

\author[A. Mortier et al.]{
A. Mortier,$^{1,2}$\thanks{E-mail: angm2@cam.ac.uk}
M.R. Zapatero Osorio,$^{3}$ 
L. Malavolta,$^{4}$ 
Y. Alibert,$^{5}$ 
K. Rice,$^{6,7}$ 
\newauthor
J. Lillo-Box,$^{3}$ 
A. Vanderburg,$^{8,\mathsection}$ 
M. Oshagh,$^{9,10}$ 
L. Buchhave,$^{11}$ 
V. Adibekyan,$^{12,13,14}$ 
\newauthor
E. Delgado Mena,$^{12,13}$ 
M. Lopez-Morales,$^{15}$ 
D. Charbonneau,$^{15}$ 
S.G. Sousa,$^{12,13}$ 
\newauthor
C. Lovis,$^{16}$ 
L. Affer,$^{17}$ 
C. Allende Prieto,$^{9,10}$ 
S.C.C. Barros,$^{12,13}$ 
S. Benatti,$^{17}$ 
\newauthor
A.S. Bonomo,$^{18}$ 
W. Boschin,$^{9,10,19}$ 
F. Bouchy,$^{16}$ 
A. Cabral,$^{20}$ 
A. Collier Cameron,$^{21}$ 
\newauthor
R. Cosentino,$^{19}$ 
S. Cristiani,$^{22}$ 
O. D. S. Demangeon,$^{12,13,14}$ 
P. Di Marcantonio,$^{22}$ 
\newauthor
V. D'Odorico,$^{22,23}$ 
X. Dumusque,$^{16}$ 
D. Ehrenreich,$^{16}$ 
P. Figueira,$^{12,24}$ 
A. Fiorenzano,$^{19}$
\newauthor
A. Ghedina,$^{19}$ 
J.I. Gonz\'alez Hern\'andez,$^{9,10}$ 
J. Haldemann,$^{5}$ 
A. Harutyunyan,$^{19}$ 
\newauthor
R.D. Haywood,$^{15,25,\mathsection}$ 
D.W. Latham,$^{15}$ 
B. Lavie,$^{16}$ 
G. Lo Curto,$^{26}$ 
J. Maldonado,$^{17}$ 
\newauthor
A. Manescau,$^{26}$ 
C.J.A.P. Martins,$^{12,13}$ 
M. Mayor,$^{16}$ 
D. M\'egevand,$^{16}$ 
A. Mehner,$^{24}$ 
\newauthor
G. Micela,$^{17}$ 
P. Molaro,$^{22,23}$ 
E. Molinari,$^{27}$ 
N.\,J. Nunes,$^{20}$ 
F.A. Pepe,$^{16}$ 
E. Palle,$^{9,10}$ 
\newauthor
D. Phillips,$^{15}$ 
G. Piotto,$^{4}$ 
M. Pinamonti,$^{18}$ 
E. Poretti,$^{19,28}$ 
M. Riva,$^{28}$ 
\newauthor
R. Rebolo,$^{9,10}$ 
N.C. Santos,$^{12,13,14}$ 
D. Sasselov,$^{15}$ 
A. Sozzetti,$^{18}$ 
A. Su\'{a}rez Mascare\~{n}o,$^{9,10}$ 
\newauthor
S. Udry,$^{16}$ 
R.G. West,$^{29}$ 
C.A. Watson,$^{30}$ 
T.\,G. Wilson$^{21}$ 
\\
\\
$^{1}$Astrophysics Group, Cavendish Laboratory, University of Cambridge, J.J. Thomson Avenue, Cambridge CB3 0HE, UK\\
$^{2}$Kavli Institute for Cosmology, University of Cambridge, Madingley Road, Cambridge CB3 0HA, UK\\
$^{3}$Centro de Astrobiolog\'\i a (CSIC-INTA), Crta. Ajalvir km 4, Torrej\'on de Ardoz, Madrid, Spain\\
$^{4}$Dip. di Fisica e Astronomia Galileo Galilei - Universit\`a di Padova, Vicolo dell'Osservatorio 2, 35122, Padova, Italy\\
$^{5}$Physics Institute, University of Bern, Sidlerstrasse 5, 3012 Bern, Switzerland\\
$^{6}$SUPA, Institute for Astronomy, Royal Observatory, University of Edinburgh, Blackford Hill, Edinburgh EH93HJ, UK\\
$^{7}$Centre for Exoplanet Science,  University of Edinburgh,  Edinburgh,  UK\\
$^{8}$Department of Astronomy, The University of Texas at Austin, 2515 Speedway, Stop C1400, Austin, TX 78712, USA\\
$^{9}$Instituto de Astrof\'{\i}sica de Canarias (IAC), Calle V\'{\i}a L\'actea s/n, 38205 La Laguna, Tenerife, Spain\\
$^{10}$Departamento de Astrof\'{\i}sica, Universidad de La Laguna (ULL), 38206 La Laguna, Tenerife, Spain\\
$^{11}$DTU Space, National Space Institute, Technical University of Denmark, Elektrovej 328, DK-2800 Kgs. Lyngby, Denmark\\
$^{12}$Instituto de Astrof\'{\i}sica e Ci\^encias do Espa\c co, Universidade do Porto, CAUP, Rua das Estrelas, 4150-762, Porto, Portugal \\
$^{13}$Centro de Astrof\'{\i}sica da Universidade do Porto, Rua das Estrelas, 
4150-762 Porto, Portugal\\
$^{14}$Departamento de F\'{\i}sica e Astronomia, Faculdade de Ci\^encias, Universidade do Porto, Rua Campo Alegre, 4169-007, Porto, Portugal\\
$^{15}$Center for Astrophysics | Harvard \& Smithsonian, 60 Garden Street, Cambridge, MA 02138 USA\\
$^{16}$Observatoire Astronomique de l'Universit\'e de Gen\`eve, Chemin des Maillettes 51, Sauverny, CH-1290, Switzerland\\
$^{17}$INAF - Osservatorio Astronomico di Palermo, Piazza del Parlamento 1, I-90134 Palermo, Italy\\
$^{18}$INAF - Osservatorio Astrofisico di Torino, via Osservatorio 20, 10025 Pino Torinese, Italy\\
$^{19}$Fundaci\'on G. Galilei -- INAF (Telescopio Nazionale Galileo), Rambla J. A. Fern\'andez P\'erez 7, 38712 Bre\~na Baja, La Palma, Spain\\
$^{20}$Instituto de Astrof\'isica e Ci\^encias do Espa\c{c}o, Faculdade de Ci\^encias da Universidade de Lisboa, Campo Grande, PT1749-016 Lisboa, Portugal\\
$^{21}$School of Physics and Astronomy, University of St Andrews, North Haugh, St Andrews, Fife, KY16 9SS, UK\\
$^{22}$INAF - Osservatorio Astronomico di Trieste, via G.B. Tiepolo 11, I-34143 Trieste, Italy\\
$^{23}$Institute for Fundamental Physics of the Universe, IFPU, Via Beirut 2, 34151 Grignano, Trieste, Italy\\
$^{24}$European Southern Observatory, Casilla, 19001, Santiago, Chile\\
$^{25}$School of Physics, University of Exeter, Stocker Road, Exeter EX4 4QL, UK\\
$^{26}$European Southern Observatory, Karl-Schwarzschild-Strasse 2, 85748, Garching b. M\"unchen, Germany\\
$^{27}$INAF - Osservatorio Astronomico di Cagliari, via della Scienza 5, I-09047, Selargius, Italy
$^{28}$INAF-Osservatorio Astronomico di Brera, Via E. Bianchi 46, 23807 Merate, Italy\\
$^{29}$Astronomy Group, Department of Physics, University of Warwick, Coventry CV4 7AL, UK\\
$^{30}$Astrophysics Research Centre, School of Mathematics and Physics, Queen's University Belfast, BT7 1NN Belfast, UK\\
$\mathsection$NASA Sagan Fellow\\
}

\date{Accepted 2020 September 28. Received 2020 September 15; in original form 2020 July 29}

\pubyear{2020}

\begin{document}
\label{firstpage}
\pagerange{\pageref{firstpage}--\pageref{lastpage}}
\maketitle

\newpage

\begin{abstract}
This paper reports on the detailed characterisation of the \thisstar\ planetary system with \emph{K2}, WASP, and ASAS-SN photometry as well as high-resolution spectroscopic data from HARPS-N and ESPRESSO. The host, \thisstar, is confirmed to be a mildly evolved ($\log g=4.17$), iron-poor ([Fe/H]$=-0.46$), but alpha-enhanced ([$\alpha$/Fe]$=0.27$), chromospherically quiet, very old thick disc G2 star. A global fit, performed by using \texttt{PyORBIT} shows that the transiting planet, \planetb, orbits with a period $P_b=5.3518\pm0.0004$\,d, and has a planet radius of $1.82^{+0.11}_{-0.09}$\,\rearth\ and a mass of $5.29^{+0.76}_{-0.77}$\,\mearth, resulting in a bulk density slightly lower than that of the Earth. The stellar chemical composition and the planet properties are consistent with \planetb\ being a terrestrial planet with an iron core mass fraction lower than the Earth. We announce the existence of a second signal in the radial velocity data that we attribute to a non-transiting planet, \planetc, with an orbital period of $15.6785\pm 0.0064$\,days, orbiting in near-3:1 mean-motion resonance with the transiting planet, and a minimum planet mass of $11.3\pm1.1$\,\mearth. Both planet signals are independently detected in the HARPS-N and ESPRESSO data when fitted separately. There are potentially more planets in this resonant system, but more well-sampled data are required to confirm their presence and physical parameters. 
\end{abstract}




\section{Introduction}

Planetary formation and evolution theories can only improve by discovering and characterising a variety of planets around a variety of stars. An early example of this was the discovery of the first exoplanet around a solar-type star, 51\,Peg\,b \citep{Mayor95}, a hot Jupiter. This type of planet challenged planet formation theories that were at the time mainly based on the structure and composition of the Solar System.

Since then, space missions like \emph{CoRoT}, \emph{Kepler}, \emph{K2}, and \emph{TESS} \citep{Mou13,Borucki10,Howell14,Ricker15} have made extraordinary progress in exploring stars with a wide range of properties including spectral type, age, location in the Galaxy, discovering thousands of planets and uncovering a wide variety of planetary systems. With radial velocity (RV) signatures of the smallest planets having semi-amplitudes of meters per second, or lower, high resolution stable spectrographs, such as HARPS \citep{Mayor03} and HARPS-N \citep{Cos12} have been essential to further characterise these systems, precisely determining planetary masses and measuring the physical and chemical properties of the host stars. Newer-generation spectrographs, like ESPRESSO \citep{Pepe14,Pepe20}, will deliver long-term RV precision of 10\,cm/s, a factor of at least 5 better than HARPS-N, and are thus expected to perform even better \citep[as shown recently in ][]{SuarezMascareno20}.

Next to precision, it is important to also ensure the accuracy of the measured parameters. In particular for the planetary mass, having well-sampled RV data over a long time span has often been needed to cover a wide range of potential periods in the systems including the orbital periods of the known transiting planets, unknown orbital periods of additional planets, and the periods that may arise from stellar surface phenomena. Inadequate sampling or a few anomalous data points have been shown to make accurate mass measurements non-straightforward and even put planet detections in doubt \citep[e.g.][]{Lop16,Raj17,Cloutier19}.

Transiting planets offer the opportunity to characterise the planet's interior composition, since both the planetary mass and radius, and thus its bulk density, can be determined. Having precisely determined elemental stellar abundances greatly helps in breaking the degeneracies between the models \citep[e.g. ][]{Dorn15,Dorn17,Hinkel18}. Studies on the Solar System and exoplanetary systems show that the stellar and planetary composition are closely linked \citep[e.g. ][]{Javoy10,San15,Santos17,Thiabaud15}. Intense follow-up of small transiting exoplanets with high-resolution spectrographs provide many high-quality spectra allowing the precise determination of stellar elemental abundances, creating the unique opportunity to study the planet's internal structure.

\thisstar, a slightly evolved G2 star, was found, with \emph{K2} data, to have a transiting planet by \citet{Fridlund17}. They confirmed the existence of \planetb, a super-Earth ($R_p = 1.9\pm0.2$\,\rearth) with a period of $5.3512$\,days. From an extensive analysis of the stellar parameters, they found that the star, originally thought to be a Hyades member, was likely to be further away than the Hyades cluster and has an age of $10.8\pm1.5$\,Gyr. It was clear from their work that a large investment of observing time to obtain RVs would be necessary to accurately and precisely characterise this system further. Less than 2\% of known exoplanets orbit a star older than 10Gyr. With \thisstar\ thus being one of the oldest planet host stars ever discovered, this system covers a yet under-explored part of the stellar host parameter space.

The size of \planetb\ makes this system even more interesting as it lies on the right-hand side of the planet radius valley found for small exoplanets \citep[e.g.][]{Ful17,VanE18,Zeng17a}. A precise planetary mass determination is thus crucial to understand whether this is a rocky planet or whether it has a large water layer or gas envelope.

The ESPRESSO and HARPS-N Science Teams have joined forces to obtain 154 precise RV observations of this system spanning over $4.5$ years. We describe in this paper how this effort resulted in a measured mass of the transiting planet \planetb, $M_p = 5.29$\,M$_\oplus$, with seven sigma significance. This is $\sim3$\,\mearth\ lower than the value determined by \citet{Fridlund17}, but compatible within errors. We furthermore claim the presence of an additional non-transiting planet, \planetc, with a minimum mass of 11\,M$_\oplus$ and a period of $15.678$\,d, making this planet orbit in near 3:1 resonance with the transiting planet \planetb. The long time coverage of our RV data rules out the existence of the more massive outer companion, hypothesised by \citet{Fridlund17}.

The paper is structured as follows. Section \ref{observations} describes the obtained data, both from photometry and spectroscopy. We describe the star, including stellar activity in Section \ref{star}. A global fit of the data is described in Section \ref{global}. We then discuss the accuracy of the mass of \planetb, the origin of the additional strong signal in the data at $15.678$\,d and potential additional signals in the data in Sections \ref{accuracy}, \ref{origin}, and \ref{extra}, respectively. Finally, we discuss this system and conclude in Section \ref{discussion}.

\section{Observations}\label{observations}

\subsection{\emph{K2} Photometry}

\thisstar\ was observed as part of the fourth Campaign of the {\it K2} mission (7th February - 23th April 2015)\footnote{Guest Observer Programmes:  GO4060\_LC, GO4033\_LC, GO4007\_LC}. It was observed in long cadence mode with integration times of $29.4$ min totaling 3168 observations over 68.5\,days.  

We obtained the data from the Mikulski Archive for Space Telescopes (MAST\footnote{\url{https://archive.stsci.edu/k2/}}). The light curve was extracted following procedures described in \citet{Vand14}, and the effects from the short-timescale spacecraft drift were corrected in a simultaneous fit of the transit shape, systematics, and low-frequency variations following \citet{Vand16a}.

Various extractions of a non-flattened \emph{K2} light curve show no convincing evidence of rotation and allows us to rule out signals with amplitudes greater than about 2\,ppt (parts-per-thousand) and periods less than about 70 days over the duration of the observation. In the remaining analyses, we only used the flattened light curve. The flattened and normalised light curve have flux uncertainties of 61 ppm (parts-per-million).

\subsection{WASP Photometry}

\thisstar\ has been observed as part of the SuperWASP programme \citep{Pollacco06}. The WASP photometric band roughly overlaps with the wavelength range of our spectroscopic data. In total, observations were taken over 5 seasons, ranging between 29th July 2004 and 29th January 2012 (i.e. spanning 7.5 years). The seasons when data were taken are 2004-5, 2006-7, 2009-10, 2010-11, and 2011-12. The light curve was extracted following procedures described in \citet{Pollacco06}. Detrending for systematic effects was done by using SysRem \citep[e.g.][]{ACC06, Mazeh07}. Stellar variability and potential transits are left in the data.

The resulting light curve was cleaned for outlier data points by performing a 5-sigma clip of the flux and removing points with unusually large errors. The final light curve, shown in Supplementary figure S1,  contained 21118 data points over 236 nights and was scaled with the median of the light curve. The median error is 8\,ppt, while the median error for the nightly binned data is 1\,ppt. This light curve is thus not precise enough to detect the transiting planet, \planetb, which has a depth of $0.2$\,ppt, but it can be used to explore stellar variability (see Section \ref{activity}).

\subsection{ASAS-SN Photometry}

We obtained public V-band and g-band photometry via ASAS-SN\footnote{Downloaded from \url{https://asas-sn.osu.edu}} \citep{Shappee14,Kochanek17}. We have V-band observations over 5 consecutive seasons, from July 2014 till November 2018, and g-band observations over 3 seasons, from September 2017 till March 2020. The photometry thus overlaps roughly with our spectroscopic data.

We removed outlier data points with a 5-sigma clip on the flux. The final light curves, shown in Supplementary figure S1, contained 1152 data points over 337 epochs for the g band and 864 data points over 320 epochs for the V-band. The median errors are 6 and 5\,ppt for the g-band and V-band, respectively. For nightly binned data, the median errors get reduced to 3\,ppt for both time series. These data are thus also not precise enough to detect the transits of small planets, but can be used to study photometric variability (see Section \ref{activity}).

\subsection{HARPS-N spectroscopy}

\thisstar\ was observed with the HARPS-N spectrograph \citep{Cos12}, which is installed at the Telescopio Nazionale Galileo (TNG) in La Palma, Spain. HARPS-N has a resolving power $R\approx115000$ and covers a wavelength range from 383 through 690\,nm. The Guaranteed Time Observation (GTO) programme of the HARPS-N Collaboration obtained 104 observations. This was complemented with 9 public observations \citep[used in ][]{Fridlund17}\footnote{Note that \citet{Fridlund17} reports twelve HARPS-N observations. However, three of those were taken with the second fiber in dark mode rather than the sky. We thus decided not to use these.}. The data were taken over three seasons, spanning 3.5yr (25th October 2015 - 24th February 2019).

All 113 observations were taken with an exposure time of 1800s and with the second fiber on the sky. The data were reduced using the standard Data Reduction Software \citep[DRS - ][]{Bara96}. A weighted cross correlation function  \citep[CCF - ][]{Pepe02b} was obtained by using a G2 mask. All CCFs were subsequently corrected for the sky background and possible Moon contamination (see \citet{Mal17} for details). RVs and the full-width-at-half-maximum (FWHM), were obtained from a Gaussian fit of the corrected CCF. As the star is relatively faint, the errors are photon-limited, with a median RV error of $2.6$\,m/s. The RVs have an RMS of $4.2$\,m/s, well above the median error.

From the 1D spectra, we also obtained the following measurements that could be an indicator of stellar activity \citep[see e.g. ][ for details]{Cincunegui07,Gom11}: H$\alpha$, NaID, Mount Wilson S-index $S_{\text{MW}}$, and $\log{R'_{HK}}$ converted from the S-index following \citet{Noy84b}. An example data table is in Table \ref{table:rvshort}. In the online supplementary material, all data are provided in Table S2 and shown in Figure S2.

To assess RV variations induced by stellar activity, we also computed chromatic RVs. These were computed by splitting the spectra in three different wavelength ranges ($3830.0-4468.9$\,\AA, $4432.6-5138.7$\,\AA, and $5103.9-6900$\,\AA)\footnote{This matches how the ESPRESSO chromatic RVs are derived with the exception of the lower blue and upper red boundary owing to HARPS-N's shorter wavelength range.}. The DRS provides the individual CCFs of each echelle order, which we corrected for Moon contamination as mentioned above. We used the central wavelength of each order to establish if the corresponding CCF was within the specified wavelength range and then coadded all the CCFs within each wavelength range. We obtained the chromatic RVs via a Gaussian fit to the resulting CCFs. We estimated the RV error by taking into account the photon noise from the total counts of the coadded CCF and the read-out noise, through the number of lines used in each order to build the CCF as returned by the DRS. We applied a correction factor to this estimate in order to closely reproduce the RV error from the full-spectrum CCF over a large range of SNR, knowing that the chromatic RVs will be inevitably noisier than the full-spectrum RVs. 

\begin{table*}
\caption{Example table of radial velocities and activity indicators for \thisstar. The full table can be found in the online supplementary material.}
\label{table:rvshort}
\begin{tabular}{llllllllllll}        
\hline\hline
Source & Time & RV & $\sigma_{\text{RV}}$ & FWHM & $\sigma_{\text{FWHM}}$ & H$\alpha$ & $\sigma_{\text{H}\alpha}$ & NaID & $\sigma_{\text{NaID}}$ & S$_{\text{MW}}$ & ... \\
 & {[}BJD-2450000] & [km\,s$^{-1}$] & [km\,s$^{-1}$] & [km\,s$^{-1}$] & [km\,s$^{-1}$] & \\
\hline
HARPS-N & $7321.7177$ & $-16.2835$ & $0.0025$ & $6.6726$ & $0.0050$ & $0.3177$ & $0.0004$ & $0.5475$ & $0.0003$ & $0.1637$ & ... \\
HARPS-N & $7323.5753$ & $-16.2785$ & $0.0024$ & $6.6673$ & $0.0049$ & $0.3220$ & $0.0005$ & $0.5553$ & $0.0003$ & $0.1732$ & ... \\
HARPS-N & $7324.6572$ & $-16.2826$ & $0.0027$ & $6.6753$ & $0.0054$ & $0.3150$ & $0.0005$ & $0.5604$ & $0.0003$ & $0.1768$ & ... \\
ESPRESSO & $8421.7397$ & $-16.4084$ & $0.0009$ & $6.8739$ & $0.0018$ & $0.3048$ & $0.0001$ & $0.5347$ & $0.0001$ & $0.1688$ & ... \\
ESPRESSO & $8424.7793$ & $-16.4061$ & $0.0009$ & $6.8763$ & $0.0017$ & $0.3055$ & $0.0001$ & $0.5314$ & $0.0001$ & $0.1700$ & ... \\
ESPRESSO & $8426.7576$ & $-16.4012$ & $0.0010$ & $6.8765$ & $0.0020$ & $0.3042$ & $0.0001$ & $0.5313$ & $0.0001$ & $0.1663$ & ... \\
... \\
\hline\hline
\end{tabular}
\end{table*}

\subsection{ESPRESSO spectroscopy}

We observed \thisstar\ with the fiber-fed Echelle Spectrograph for Rocky Exoplanets and Stable Spectroscopic Observations \citep[ESPRESSO -][Pepe et al. 2020 submitted]{Pepe14} installed at the incoherent combined Coud\'e facility of the Very Large Telescope (VLT) on the Paranal Observatory (Chile). A total of 41 ESPRESSO spectra were obtained as part of the GTO programmes 1102.C-0744, 1102.C-0958, and 1104.C-0350 between 30th October 2018 and 3rd March 2020. There was a technical intervention on the instrument in the second half of June 2019. This technical intervention led to a shift of the RV zero-point, and as a consequence the two data sets have to be treated independently. We refer to the 16 and 25 spectra taken before and after this intervention as ESPRESSO 1 and ESPRESSO 2, respectively.

The typical exposure time per observation was 900\,s and all data were taken with the HR21 mode (fiber size of 1\,arcsec and 2$\times$1 binning on the detector), the second fiber on the sky, and one of the VLT Telescopes. The airmass of all observations was always lower than 2.2 to guarantee a good correction of the atmospheric dispersion.

ESPRESSO raw data were reduced with the Data Reduction Software pipeline version 2.2.1, which included bias subtraction, correction for hot pixels, cosmic ray hits and flat-field, optimal extraction of the echelle orders, sky subtraction, wavelength calibration using the combined Fabry-Perot---ThAr solution, and blaze correction. The ESPRESSO data have a mean resolving power $R \approx$ 138,000, cover wavelengths from 380 through 788 nm (going redder than HARPS-N), and have typical signal-to-noise ratio (S/N) of 45--80 at 550 nm (with few exceptions, which have lower S/N values).

We used the ESPRESSO sky-subtracted spectra to measure RVs. The pipeline provides a cross-correlation function (CCF) for each ESPRESSO spectrum, which was built by using a G9 mask, an RV step of 0.5 km\,s$^{-1}$, and a systemic RV of $-16.66$ km\,s$^{-1}$. Similarly to HARPS-N data, ESPRESSO RVs and FWHMs were obtained from the Gaussian fit of the CCFs. We also obtained the same stellar-activity spectral indices from the sky-subtracted merged spectra as we did for the HARPS-N spectra. The final RVs have an RMS of $3.7$\,m/s, well above the median RV error of $1.1$\,m/s. The ESPRESSO RV errors are, like the HARPS-N RVs photon-noise-limited. An example data table is in Table \ref{table:rvshort}. In the online supplementary material, all data are provided in Table S2 and shown in Figure S2.

Chromatic RVs (blue, green, and red) were also obtained with similar wavelength ranges as the HARPS-N chromatic RVs (except a longer range for the red RVS thanks to the extended wavelength range of ESPRESSO). Details on the extraction can be found in \citet{SuarezMascareno20}.

The HARPS-N and ESPRESSO 1 RV time series overlap between October 2018 and February 2019. This is convenient for defining a proper zero point for the two data sets and finding a robust solution for the planetary system. The ESPRESSO RVs before and after the technical intervention on the spectrograph were treated as sets of data coming from two different instruments in our analysis.

\section{Stellar characterisation}\label{star}

In this section we measure and refine all the parameters of the G2 star, \thisstar, using new Gaia astrometry, and our obtained photometric and spectroscopic data. All updated stellar property values are summarised in Tables \ref{table:star} and \ref{table:star2}.

\begin{table}
\caption{\thisstar\ stellar properties}            
\label{table:star}
\begin{tabular}{l l l}        
\hline\hline
Parameter & Value & Source \\
\hline
\multicolumn{3}{l}{\emph{Designations and coordinates}} \\
\emph{K2} ID & 111 & \\
EPIC ID & 210894022 & \\
2-MASS ID & J03593351+2117552 & \\
Gaia DR2 ID & 53006669599267328 & \\
RA (J2000) & 03:59:33.54 & 1\\
Dec (J2000) & +21:17:55.24 & 1\\
\hline
\multicolumn{3}{l}{\emph{Magnitudes and astrometric solution}}\\
B & $11.80 \pm 0.03$ & 2 \\
V & $11.14 \pm 0.04$ & 2 \\
G & $10.9294 \pm 0.0006$ & 1 \\
g & $11.44 \pm 0.04$ & 2 \\
r & $10.95 \pm 0.02$ & 2 \\
J & $9.77 \pm 0.02$ & 3 \\
H & $9.48 \pm 0.03$ & 3 \\
K & $9.38 \pm 0.02$ & 3 \\
W1 & $9.32 \pm 0.02$ & 4 \\
W2 & $9.35 \pm 0.02$ & 4 \\
W3 & $9.21 \pm 0.03$ & 4 \\
Parallax $\pi$ [mas] & $4.9626 \pm 0.0674$ & 1\\
Distance $d$ [pc] & $ 201.7 \pm 2.7 $ & 5 \\
$\mu_\alpha$ [mas/yr] & $122.337 \pm 0.182$ & 1 \\
$\mu_\delta$ [mas/yr] & $-35.438 \pm 0.093$ & 1 \\
 U [km/s] & $-24.80\pm0.90$ & 5 \\
 V [km/s] & $-103.96\pm1.11$ & 5 \\
 W [km/s] & $60.48\pm1.12$ & 5 \\
\hline
\hline
\end{tabular}
\newline \centering 
1: Gaia DR2 \citep{Gaia2}; 2: UCAC-4 \citep{Zacharias12}; 3: 2MASS \citep{Cutri03}; 4: WISE \citep{Cutri13}; 5: this work
\end{table}

\begin{table}
\caption{\thisstar\ stellar properties}            
\label{table:star2}
\begin{tabular}{l l l}        
\hline\hline
Parameter & Value & Source \\
\hline
\multicolumn{3}{l}{\emph{Atmospheric parameters}}\\
$T_{\rm{eff}}$ [K] & $5775 \pm 60$ & 1, 2 \\
log\,$g$ [cgs] &  $4.25 \pm 0.15$ & 1, 2 \\
log\,$g$ [cgs] &  $4.17 \pm 0.01$ & 1, 5 \\
$[\rm{Fe/H}]$ [dex] & $-0.46  \pm 0.05$ & 1, 2 \\
$\xi_t$ [km/s] & $1.02 \pm 0.05$ & 1, 3 \\
v$\sin$\,i [km/s] & $1.1 \pm 0.5$ & 1, 4 \\
\hline
\multicolumn{3}{l}{\emph{Elemental abundances}}\\
$[\rm{C/H}]$ [dex] & $-0.30\pm0.06$ & 1, 2 \\
$[\rm{O/H}]$ [dex] & $0.03\pm0.07$ & 1, 2 \\
$[\rm{Na/H}]$ [dex] & $-0.38\pm0.02$ & 1, 2 \\
$[\rm{Mg/H}]$ [dex] & $-0.14\pm0.06$ & 1, 2 \\
$[\rm{Al/H}]$ [dex] & $-0.22\pm0.01$ & 1, 2 \\
$[\rm{Si/H}]$ [dex] & $-0.27\pm0.04$ & 1, 2 \\
$[\rm{S/H}]$ [dex] & $-0.2\pm0.07$ & 1, 2 \\
$[\rm{Ca/H}]$ [dex] & $-0.24\pm0.03$ & 1, 2 \\
$[\rm{Sc/H}]$ [dex] & $-0.23\pm0.09$ & 1, 2 \\
$[\rm{Ti/H}]$ [dex] & $-0.16\pm0.03$ & 1, 2 \\
$[\rm{V/H}]$ [dex] & $-0.28\pm0.04$ & 1, 2 \\
$[\rm{Cr/H}]$ [dex] & $-0.43\pm0.04$ & 1, 2 \\
$[\rm{Mn/H}]$ [dex] & $-0.64\pm0.05$ & 1, 2 \\
$[\rm{Co/H}]$ [dex] & $-0.33\pm0.03$ & 1, 2 \\
$[\rm{Ni/H}]$ [dex] & $-0.41\pm0.02$ & 1, 2 \\
$[\rm{Cu/H}]$ [dex] & $-0.46\pm0.07$ & 1, 2 \\
$[\rm{Zn/H}]$ [dex] & $-0.26\pm0.05$ & 1, 2 \\
$[\rm{Sr/H}]$ [dex] & $-0.53\pm0.08$ & 1, 2 \\
$[\rm{Y/H}]$ [dex] & $-0.53\pm0.06$ & 1, 2 \\
$[\rm{Zr/H}]$ [dex] & $-0.42\pm0.05$ & 1, 2 \\
$[\rm{Ba/H}]$ [dex] & $-0.59\pm0.06$ & 1, 2 \\
$[\rm{Ce/H}]$ [dex] & $-0.60\pm0.09$ & 1, 2 \\
$[\rm{Nd/H}]$ [dex] & $-0.44\pm0.04$ & 1, 2 \\
$[\alpha\rm{/Fe}]$ [dex] & $0.27$ & 5 \\
\hline
\multicolumn{3}{l}{\emph{Mass, radius, density, age}} \\
M$_*$ [\msun] &  $0.84 \pm 0.02$ & 1, 2, 6 \\
R$_*$ [\rsun] & $1.25 \pm 0.02$ & 1, 2, 6 \\
$\rho_{*}$ [$\rho_\odot$] & $0.43 \pm 0.02$ & 1, 2, 6 \\
$\rho_{*}$ [g\,cm$^{-3}$] & $0.60 \pm 0.03$ & 1, 2, 6 \\
Age $t$ [Gyr]  & $13.5^{+0.4}_{-0.9}$ & 1, 2, 6 \\
Age $t$ [Gyr]  & $12.3\pm0.7$ & 1, 2, 7 \\
\hline\hline
\end{tabular}
\newline \centering 
1: this work; 2: Adopted averaged parameters; 3: From ARES/MOOG analysis; 4: From SPC analysis; 5: using Mg, Si, Ti as the alpha abundance; 6: From isochrones analysis; 7: From chemical clocks
\end{table}

\subsection{Astrometry}

Prior to the Gaia mission \citep{Gaia16}, this star was thought to be part of the Hyades cluster of young stars \citep{Pels75}. However, with an improved new parallax and thus stellar distance, we now know \thisstar\ \citep[$200.4 \pm 2.7$\,pc -][]{Bailer-Jones18} is located behind the Hyades \citep[$47.50\pm0.15$\,pc - ][]{Bailer-Jones18} and is not part of it. 

The GaiaDR2 RUWE (renormalised unit weight error) parameter, an indicator of the quality of the astrometric solution, is $1.22$ which is towards the higher end (with RUWE peaking at 1.0 and RUWE $>1.4$ considered bad solutions). There is thus some evidence the astrometric solution is noisy. Recent work has shown that this may indicate a hint of a more massive companion \citep{Belokurov20}. Our RVs rule out a massive non-inclined companion, but a very inclined long-period massive companion could go undetected. Gaia DR3 will be required to shed more light on this.

\citet{Fridlund17} derived Galactic space velocities and determined the star is part of the thick disc population. We rederived these velocities using the Gaia DR2 data \citep{Gaia2}. We used the RV value from Gaia Dr2 (RV = $-16.66 \pm 0.72$ km/s) because Gaia has contrasted RV zero points. We also employed the trigonometric parallax and proper motions listed in Table\,\ref{table:star} to calculate the $U$, $V$, and $W$ heliocentric velocity components in the directions of the Galactic center, Galactic rotation, and north Galactic pole, respectively, with the formulation developed by \citet{johnson87}. Note that the right-handed system is used and that we did not subtract the solar motion from our calculations. The uncertainties associated with each space velocity component were obtained from the observational quantities and their error bars after the prescription of \citet{johnson87}. 

The large $V$ and $W$ components are a clear signpost that \thisstar\ kinematically belongs to the old population of the Galaxy, which agrees with the derived stellar age. More precisely, using the work from \citet{Reddy06}, and a Monte Carlo approach using 1 million samples, this star has a $97.41\pm0.06\%$ probability of belonging to the thick disc, $2.35\pm0.11\%$ probability of being a halo star and $0.25\pm0.06\%$ probability that it belongs to the thin disc. This high probability that \thisstar\ belongs to the thick disc agrees with the conclusion from \citet{Fridlund17}.

\subsection{Stellar atmospheric parameters}

We obtained stellar atmospheric parameters from the high-resolution HARPS-N spectra via three independent methods: ARES+MOOG\footnote{ARESv2: \url{http://www.astro.up.pt/~sousasag/ares/}; MOOG 2017: \url{http://www.as.utexas.edu/~chris/moog.html}}, \texttt{CCFPams}\footnote{\url{https://github.com/LucaMalavolta/CCFpams}}, and SPC. Our final adopted parameters, listed in Table \ref{table:star2}, are the average of the results from these three methods.

The ARES+MOOG method is explained in detail in \citet{Sou14} and references therein. It is a curve-of-growth method based on neutral and ionised iron lines. Equivalent widths of these spectral lines were measured automatically from the stacked HARPS-N spectrum using \texttt{ARESv2} \citep{Sou15}. Effective temperature $T_{\text{eff}}$, surface gravity $\log g$, iron abundance [Fe/H], and microturbulent velocity $\xi_t$ are then determined by imposing excitation and ionisation equilibrium. For this purpose, we used the radiative transfer code \texttt{MOOG} \citep{Sne73}, assuming local thermodynamic equilibrium (LTE) and employing a grid of ATLAS plane-parallel model atmospheres \citep{Kur93}. Subsequently, the surface gravity was corrected for accuracy \citep{ME14} and we added systematic errors in quadrature to our precision errors for the effective temperature, surface gravity, and iron abundance \citep{Sou11a}. We obtained $T_{\text{eff}}=5794 \pm 66$\,K, $\log g = 4.35 \pm 0.11$, [Fe/H]$=-0.44 \pm 0.04$, and $\xi_t = 1.02 \pm 0.05$\,km/s.

The \texttt{CCFpams} method uses the equivalent width of CCFs to obtain the effective temperature, surface gravity and iron abundance via an empirical calibration. More detail on this method can be found in \citet{Malavolta17b}. Like the ARES+MOOG method, we subsequently corrected the surface gravity for accuracy. From this method, we obtained $T_{\text{eff}}=5762 \pm 34$\,K, $\log g = 4.21 \pm 0.22$, and [Fe/H]$=-0.48 \pm 0.03$ (internal errors only).

Finally, the Stellar Parameter Classification tool (SPC) was used on the individual HARPS-N spectra. SPC is a spectrum synthesis method and is described in detail in \citet{Buch12} and \citet{Buch14}. Final values were determined by the weighted average of the individual results, where the signal-to-noise ratio was used as the weight. Due to known issues related with the spectroscopic determination of the surface gravity \citep{Tor12,ME14}, we constrained this parameter by using isochrones. We obtained $T_{\text{eff}}=5769 \pm 49$\,K, $\log g = 4.19 \pm 0.10$, [m/H]$=-0.46 \pm 0.08$, and projected rotational velocity $v\sin i = 1.1 \pm 0.5$\,km/s. SPC determines global metallicity assuming solar relative abundances for all the metals, while \texttt{CCFPams} and ARES+MOOG determine specifically iron abundance. Due to the overwhelming amount of iron lines in the HARPS-N wavelength range, these measures are treated as the same in this work.

We additionally analysed the stacked ESPRESSO spectrum with ARES+MOOG. The results were in full agreement with the results from the HARPS-N spectra: $T_{\text{eff}}=5779 \pm 62$\,K, $\log g = 4.37 \pm 0.10$, [Fe/H]$=-0.47 \pm 0.04$, and $\xi_t = 1.06 \pm 0.03$\,km/s. The two other methods (CCFPams and SPC) would require significant changes to the code and additional calibrated measurements in order to be applied to the ESPRESSO spectrum. We hence decided to adopt the averaged parameters from the HARPS-N spectra only.

The derived parameters show that \thisstar\ is mildly evolved, with a solar-like temperature but much more iron-poor than the Sun.

\subsection{Stellar abundances}

We used the stacked HARPS-N and stacked ESPRESSO spectra to derive elemental abundances for the following elements: C, O, S, Na, Mg, Al, Si, Ca, Ti, Cr, Ni, Co, Sc, Mn, V, Cu, Zn, Sr, Y, Zr, Ba, Ce, and Nd. We used the same setup as for the ARES+MOOG method described above and ran the analysis for these elements also in LTE, like we did for the atmospheric parameters. The methods are described in more detail in \citet{Adi12,ME13c,BertranLis15,DelgadoMena17}. All values are relative to the Sun, with the values of \citet{Asplund09} as a reference for the solar values. In Table \ref{table:star2}, we have averaged the individual values from the HARPS-N and ESPRESSO spectra.

We find that \thisstar\ is alpha-enhanced with $[\alpha/$Fe$] = 0.27$ (using the average of the magnesium, silicon, and titanium abundances as the alpha-abundance). This is in line with the star belonging to the thick disc following \citep{Adi12}. This is also confirmed by the enhanced [O/Fe] and [Zn/Fe]. The overall enhancement with respect to iron for these elements is in agreement with iron-poor planet hosts being generally enhanced more in alpha-elements \citep{Haywood09,Adi12b}.

\subsection{Stellar mass, radius, age, and distance}

Stellar parameters were obtained from isochrones and evolutionary tracks. As input, we used the spectroscopically determined effective temperature and iron abundance, the Gaia DR2 parallax, and 11 magnitudes (listed in Table \ref{table:star}) ranging from the visible to the mid-infrared. 

We used the code \texttt{isochrones} \citep{Morton15}, which uses \texttt{MultiNest} \citep{Feroz19} for its likelihood analysis. We ran the code three times, varying only between the three sets of spectroscopic parameters. We chose 400 live points, stellar models from the Dartmouth Stellar Evolution Database \citep{Dotter08} and constrained the prior on the age to be lower than $\log$(Age/yr)$=10.15$. We extracted the final values and errors from the combined posteriors of these three runs, taking the median and the 16th and 84th percentile. These are the values reported in Table \ref{table:star2}. The results of each individual run are all within one sigma of each other, owing to the agreement between the spectroscopic parameters.

We additionally tried obtaining parameters by using the models from the MESA isochrones and Stellar Tracks \citep[MIST - ][]{Dotter16}. However, the MultiNest evidences decisively favoured the Dartmouth models (with Bayes factors exceeding $10^9$). Looking into this, we found that the MIST models as used by the \texttt{isochrones} package do not always interpolate well for stars older than 10\,Gyr often giving back nan values. Since \thisstar\ is clearly older, it fails to find a good solution via the MIST models.

The stellar distance obtained via this method, $201.7\pm2.7$\,pc, agrees very well with the one obtained by \citet{Bailer-Jones18}, $200.4 \pm 2.7$\,pc. The surface gravity as calculated from the stellar radius and mass, $\log g = 4.17\pm0.01$ is much more precise than the spectroscopically determined surface gravity, but they do agree within 1 sigma. It confirms the star is mildly evolved.

We found a stellar mass of $M_\ast=0.84\pm0.02$\msun\ and a stellar radius of $R_\ast=1.25\pm0.02$\rsun. \citet{Fridlund17} determined the stellar mass and radius via 3 different methods. Our results agree within 1 sigma with two of their results (the DSEP and PARAM results). However, we do not find agreement with the stellar mass and radius they used to calculate their planetary mass and radius which were higher ($1.0\pm0.07$\,M$_\odot$ and $1.4\pm0.14$\,R$_\odot$, respectively) than the values we obtained. This resulted in a systematic overestimation of the planetary mass and radius they reported as compared to ours.

From the isochrones analysis, we find that the age is $13.5^{+0.4}_{-0.9}$\,Gyr. This is an older and more precise stellar age than the one reported by \citet{Fridlund17}, who reported $10.8\pm1.5$\,Gyr. Stellar evolution models are, however, not designed for stars this old. We additionally made use of the stellar chemical abundances to calculate ages by using the so-called chemical clocks (i.e. certain chemical abundance ratios which have a strong correlation for age). We applied the 3D formulas described in Table 10 of \citet{DelgadoMena19}, which also consider the variation in age produced by the effective temperature and iron abundance [Fe/H]. The chemical clocks [Y/Mg], [Y/Zn], [Y/Ti], [Sr/Ti], and [Sr/Zn] were used from which we took a weighted average. We derive an age of $12.3\pm 0.7$\,Gyr. This is lower than the isochrone-constrained age, but still points to a very old star. From the study of \citet{Nissen20}, it is also clear that the chemical composition of \thisstar\ is consistent with the star being very old. While we thus may not accurately determine the age of this system, we do show that it is very likely older than 10\,Gyr. This agrees with the kinematic behaviour of this star pointing to a thick disc.

\subsection{Stellar activity}\label{activity}

Signals arising from stellar activity are the main barrier in detecting small exoplanets in RVs and accurately and precisely measuring their mass \citep[e.g. ][]{Hay14,Raj15,ME16,Dum17}. It is thus paramount to understand the star, its rotation period, and the strength of possible activity effects to the best of our ability. We have used both photometric and spectroscopic data for this purpose.

\subsubsection{From photometry}

The WASP data were taken years before the first RVs were observed. However, given its 7.5yr data span, it could give us insight into the stellar rotation period and the overall photometric variability. The RMS of the full light curve is equal to the median error, 8ppt. Split into the five seasons, we find that the third season was marginally the most quiet with an RMS of 6ppt while the first season had the largest RMS of 10ppt. Using the nightly binned data reduced the RMS to 1-2 ppt, again equal to the median error. Any stellar variability during the time span of the data is thus likely lower than or at the level of the data errors.

\begin{figure*}
    \includegraphics[width=\textwidth]{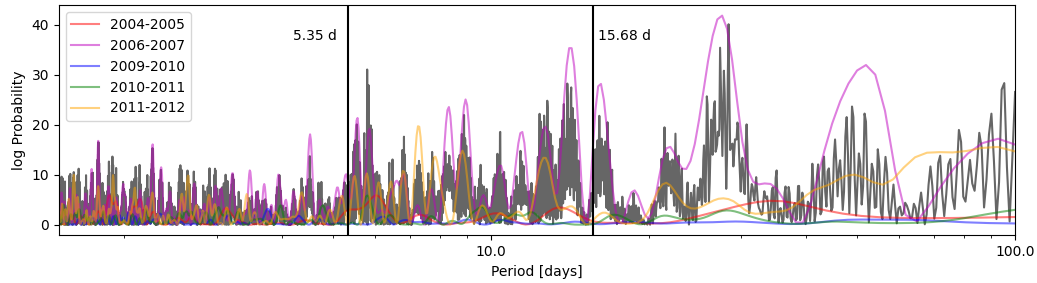}
    \caption{BGLS periodogram for the full WASP data (black curve) and the individual seasons (coloured curves). The two vertical lines indicate the periods of the transiting planet and the strongest RV signal at 15.678\,d.}
    \label{fig:wasp}
\end{figure*}

We applied a Bayesian General Lomb-Scargle periodogram \citep[BGLS - ][]{ME15} to the full WASP light curve and to the five seasons separately. This is shown in Figure \ref{fig:wasp}. Using all data or nightly binned data does not affect the shape of the periodograms. For most seasons, no strongly significant periodic signals are evident, with the exception of season two (2006-2007). This season contains the most data points and shows its strongest peaks at 27.5, 14, and 52 days. These could be related to the rotation of the star. We caution, however, that the strength of the $27.5$\,d periodicity may be partly sampling-related due to avoiding Moon-lit nights. Indeed, when we phasefold the data in season two with a period of 27.5 days, an obvious phase gap is present coinciding with a full Moon.

The BGLS periodogram of the full data shows forests of peaks at 28, 14 and 50 days, all coinciding with the BGLS periodogram from season 2. Additionally, the BGLS periodogram of the full data shows periodicity around 5.8\,days. Given the age, evolutionary status, and projected rotational velocity of \thisstar\ (see Section \ref{star}), it is highly unlikely this is related to the rotation period. However, it is important to bear in mind that this photometric variability is present, given its period is close to the orbital period of the transiting planet, \planetb.

As a second investigation, we applied a Gaussian Process (GP) regression with a quasi-periodic kernel to the full WASP data. We used the nightly binned data, normalised with the median of the data, since that speeds up the process considerably and the periodograms of binned and unbinned data look similar. We used a quasi-periodic covariance kernel \citep[as described in e.g. ][]{Grunblatt15,Dubber19} and allowed for additional white noise. The \texttt{PyORBIT} code \citep{MalavoltaPyorbit} was used to do the analysis and \texttt{MultiNest} was used for the parameter inference, with 400 live points. We constrained the rotation period to be lower than 100\,d, the decay timescale to be lower than 1000\,d, and the coherence scale to be between $0.2$ and $2.0$, shown by several works \citep[e.g.][]{Hay14,Dubber19} to be the most physically motivated for stellar activity phenomena\footnote{Not constraining the coherence scale led to multimodal posterior distributions and a coherence scale of $0.08$}. Taking the median, 16th and 84th percentile of the posterior distributions, the results are as follows: rotation period $P_{\text{rot}}= 27.4^{+1.4}_{-1.2}$\,d, decay timescale $P_{\text{dec}} =83^{+255}_{-51}$\,d, amplitude $h=1.2\pm0.3$\,ppt, coherence scale $w=0.32^{+0.32}_{-0.09}$, and white noise $s=1.2\pm0.2$\,ppt. This model is clearly preferred over a white-noise-only model with a Bayes factor\footnote{We use the Nested Importance Sampling global evidences to compute Bayes factors throughout this work \citep{Feroz19}} exceeding $10^5$.

\begin{figure*}
    \includegraphics[width=\textwidth]{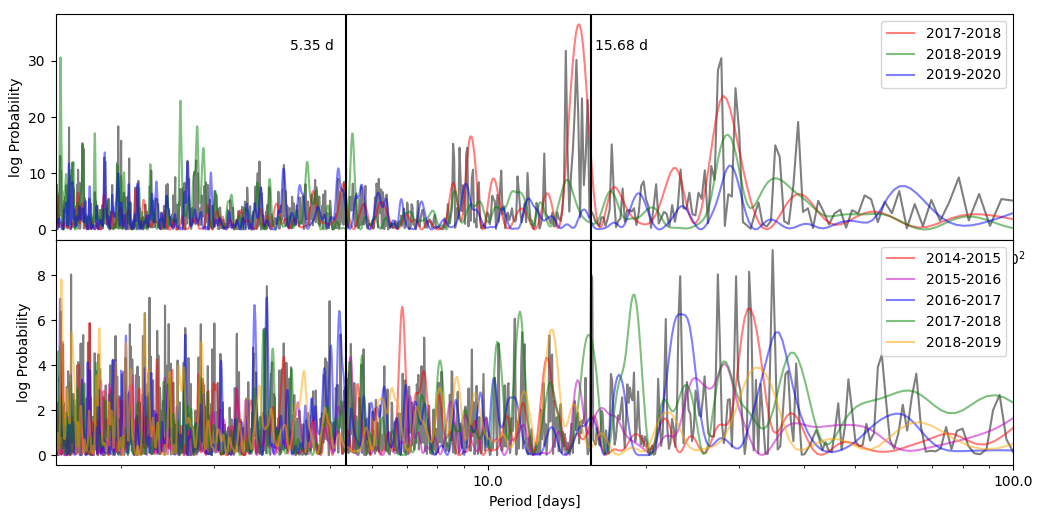}
    \caption{BGLS periodograms for the full ASAS-SN data (black curve) and the individual seasons (coloured curves). Top plot is for g-band data and bottom plot for V-band data. The two vertical lines indicate the periods of the transiting planet and the strongest RV signal at 15.678\,d.}
    \label{fig:asas}
\end{figure*}

We repeated the same process for the ASAS-SN data. The RMS of the full light curve is 13 and 8\,ppt, for g-band and V-band, respectively. A higher RMS for blue with respect to red wavelengths is expected if the star is faculae-dominated instead of spot-dominated. Using the binned data reduces the RMS to 10 and 6\,ppt, respectively. Both these values are higher than the median error and higher than the results from the WASP light curve. Since the WASP and ASAS-SN data are not concurrent, with at least two years between the end of the WASP data and the start of the ASAS-SN data, this could suggest that \thisstar\ was in a quieter part of its activity cycle when the WASP data were taken and slightly more active at the time of the RV observations or it could be a simple consequence of the WASP data being more precise. In any case, overall these RMS values are still fairly low. Using the binned data, we find that the V-band data are reasonably constant in RMS. The RMS of the seasonal g-band data increases with each season. It is noteworthy that this is in contrast with the average S index going down over the same seasons as shown in the next Section.

The BGLS periodogram of the ASAS-SN data is shown in Figure \ref{fig:asas}. As expected from the RMS values, the g-band data shows stronger periodicities than the V-band data with the latter not showing any significant periodicities. For the g-band data, the first season shows the strongest periodicities at 14.9, 28 and 9 days, similar to the WASP data. We do still have to caution that this variability may be partly related to the cycle of the Moon. When the first season of data is phased up with the Moon phases, the quadratures do appear to be close to full and new Moon. The second season also identifies some periodicity at 5.5\,d, slightly higher than the period of \planetb.

For the ASAS-SN binned data, we also ran a GP regression, in a similar fashion as described above. For both datasets, the model with GP was decisively preferred over a model with white noise only. The results are as follows for the g band data: $P_{\text{rot}}= 28.1^{+1.2}_{-0.2}$\,d, $P_{\text{dec}} =571^{+312}_{-387}$\,d, $h=11.9^{+8.6}_{-4.6}$\,ppt, $w=0.94^{+0.55}_{-0.32}$, and white noise $s=8.0\pm0.4$\,ppt. The value for the rotation period broadly agrees with the results from the WASP data but all other parameters are larger. The large amplitude is worrisome given that the \emph{K2} photometry firmly ruled these amplitudes out. The g-band is bluer than the Kepler band, so stellar signals could be expected to be somewhat larger there, but it could also be that this is a spurious signal, not related to the rotation of the star. The V-band data gave $P_{\text{rot}}= 31.2^{+5.8}_{-7.9}$\,d, $P_{\text{dec}} =105^{+70}_{-51}$\,d, $h=2.9^{+0.8}_{-0.6}$\,ppt, $w=0.83^{+0.69}_{-0.38}$, and white noise $s=4.3\pm0.4$\,ppt, in better agreement overall with the WASP data, but still larger than what we would expect given the \emph{K2} data.

In total, all our photometry spans almost 15 years. A GP regression was thus also run on all three photometry data series simultaneously. The rotation period, decay timescale, and coherence scale were kept the same across all three data series as these values are inherent to the star. The amplitude and white noise could vary per data set. Since each data set was taken in a different photometric band, we would not expect the amplitude to be the same across the data sets. The results are $P_{\text{rot}}= 29.2\pm0.8$\,d, $P_{\text{dec}} =74^{+48}_{-23}$\,d, and $w=0.59^{+0.17}_{-0.12}$. The amplitudes for the WASP, ASAS-SN g-band, and ASAS-SN V-band data, respectively are $1.2\pm0.3$, $6.0^{+1.5}_{-1.0}$, and $2.8\pm0.6$\,ppt, with additional white noise of $s=1.3\pm0.1$, $7.8\pm0.4$, and $4.2\pm0.3$\,ppt.

From the sheer length of the complete data set it seems plausible from the WASP and ASAS-SN photometry that the stellar rotation period of \thisstar\ is indeed around 29 days, but with relatively low amplitude. However, the \emph{K2} photometry, even though shorter in time span, does not corroborate such a signal. We injected signals with a period of 27 days into the \emph{K2} light curve with varying amplitudes. Based on a visual inspection of the results, we expect that if such a signal was present in the \emph{K2} light curve with a semi-amplitude greater than about 500\,ppm, we likely would have detected it. There is thus still a chance the variations seen in the other photometry are influenced by Moon light and thus exaggerated. We estimated the expected RV RMS based on the photometry variations we see using the results of \citet{Hojjatpanah20} which are based on TESS photometry. While there is large scatter in their relations and considering the TESS passband is redder than the photometry used here, we cannot exclude 1-2\,m/s RMS due to stellar activity in our data.

\subsubsection{From spectral indicators}

Next, we investigated any effects from stellar activity seen in the spectra and its derivative measurements. For this we used the measured projected rotational velocity and five different activity indicators, namely FWHM, H$\alpha$, NaID, $S_{\text{MW}}$, and $\log{R'_{HK}}$.

A rough constraint on the rotational period can be set by using the projected rotational velocity and the stellar radius, and assuming a stellar inclination of 90 degrees. Using the values in Table \ref{table:star2}, we get $P_{\text{rot}}= 57^{+48}_{-18}$\,days, from drawing 10000 samples. Due to degeneracies between the various broadening mechanisms given our spectral resolution, it is only possible to define an upper limit of $2$\,km/s for stars rotating slower than that. If we thus use the latter value as an upper boundary, we find that the rotational period should be higher than $31.6\pm0.5$\,days. While both these values are higher than the 29\,d period derived from the photometry, it is worth bearing in mind that the stellar inclination is likely not exactly 90 degrees and the actual rotation periods are thus likely a bit lower. A rotation period of 29\,d is thus not excluded from the projected rotational velocity and stellar radius.

\begin{figure*}
    \includegraphics[width=\textwidth]{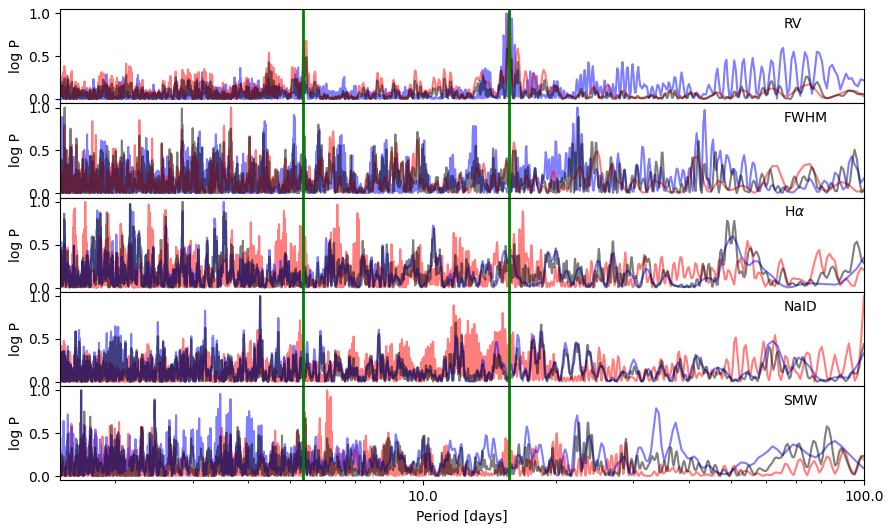}
    \caption{BGLS periodograms for, from top to bottom, RV, FWHM, H$\alpha$, NaID, and S-index. The blue curve is for HARPS-N data, the red curve for ESPRESSO data and the black curve for the combined data with the appropriate offsets applied where necessary. The green vertical lines indicate the transiting planet period, $5.35$\,d, and the strongest periodicity seen in the RVs, $15.68$\,d. The y-axis, showing the period posterior probability,  has been normalised to be between 0 and 1.}
    \label{fig:BGLS}
\end{figure*}

We also investigated the stellar activity indicators derived from the spectra. For this investigation, we subtracted three offsets from the RVs, for the HARPS-N, ESPRESSO 1, and ESPRESSO 2 data, respectively. These offsets were derived from the final global fits: $-16.27535$, $-16.40504$, and $-16.40844$\,km/s, respectively. We also subtracted the median value of FWHM for each of these three data sets since the FWHM of the CCF is instrument- and pipeline-dependent. The other indicators should be independent from the instrument so no values were subtracted for those indicators. We note that due to some interstellar clouds between us and \thisstar, and the corresponding extra Na absorption lines (shown in \citet{Fridlund17} and confirmed in our spectra), the NaID index may be contaminated.

Figure \ref{fig:BGLS} shows the BGLS periodogram for the RVs as well as the indicators. The BGLS periodogram for $\log{R'_{HK}}$ has the same shape as the S-index and is not shown. The periodograms have been normalised to put the maximum at 1. The periodograms of the RVs for both instruments shows its highest peak at 15.678\,d. The ESPRESSO RVs show additional peaks at the period of \planetb, 5.35\,d, and also at 4.47\,d. The HARPS-N RVs show marginal power at 5.35\,d but some more peaks at higher periods, with forests of peaks around 23\,d, 30\,d, and higher than 50\,d.

In contrast with the RV periodogram, the periodograms of the indicators are not as clear and no strong periodicities stand out. Crucially, none of the indicators show any power around 29\,d, casting doubt on the origin of the 29\,d signal in the photometry. The HARPS-N FWHM displays periodicities around 43, 22, and 13\,d, suggesting the rotation period could perhaps be as high as 43\,d where the other periodicities would be the harmonics. None of these periodicities are strong enough to draw a firm conclusion and they are not replicated in the other indicators either. The S-index shows different behaviour for each instrument, with marginally likely periodicities at 6\,d for ESPRESSO and 34\,d for HARPS-N. Overall, the periodicities in the activity indicators (or lack thereof) suggest that \thisstar\ is a quiet star. This is not unexpected given its old age.

The S-index and associated $\log{R'_{HK}}$ are traditionally seen as an excellent indicator for a star's magnetic cycle. The Sun for example has an S-index varying between $0.16$ and $0.18$ throughout its 11-year magnetic cycle \citep[e.g. ][]{Egeland17}. There is a hint that the S-index of \thisstar\ goes down over the course of the RV observations, with a median S-index going from $0.171$ in the first season of data to $0.165$ in the final season of data. The median $\log{R'_{HK}}$ goes from $-4.91$ to $-4.94$ with an overall median value of $-4.93$. Using the calibrations from \citet{Noy84} and \citet{Mama08}, this median value of $\log{R'_{HK}}$ would indicate a rotation period of $25$\,d, but these calibrations are not well tested for stars as old or alpha-enhanced as \thisstar.

The average value of $\log{R'_{HK}}$ can be used to estimate the expected stellar-induced RV variations. From equation 1 of \citet{Hojjatpanah20}, a value of $-4.93$ for $\log{R'_{HK}}$ translates to an RV RMS of $2.85$\,m/s. Many stars included in their fit were, however, rotating much faster than \thisstar. In contrast, \citet{SuarezMascareno17} estimates that the RV semi-amplitude induced by stellar activity variations of a G dwarf with average $\log{R'_{HK}}$ of $-4.9$ should be lower than 1\,m/s, though this was based on a smaller sample. Recent HARPS-N solar data shows an RMS of $1.63$\,m/s \citep{ACC19} while the Sun was approaching Solar minimum with values of $\log{R'_{HK}}$ around $-4.97$. We can thus reasonably expect RV variations from stellar activity at the level of $0.5 - 3$\,m/s.

Due to the lack of common periodicities between the RVs and the spectral indicators, it is no surprise that there is little to no correlation between these values. Pearson and Spearman correlation coefficients of the RVs and indicators are all between $-0.2$ and $0.2$ indicating no significant correlation between RVs and the indicators.

To conclude, \thisstar\ appears to be a chromospherically quiet star. The rotation period can not be uniquely determined from the plethora of data and indicators. The photometry points marginally to a rotation period of 29\,d, while the spectroscopic indicators point marginally to 25 or 43 days and do not confirm clear variability at 29d. The RVs show no sign of strong periodicities at any of these periods.

\section{Global fit}\label{global}

\begin{figure*}
    \includegraphics[width=\textwidth]{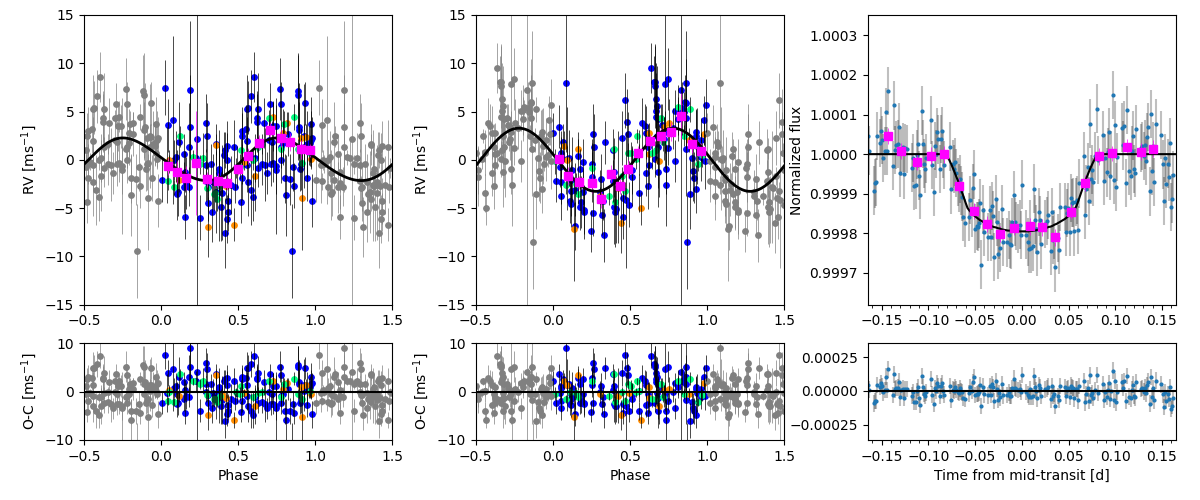}
    \caption{Solution from the global fit two-Keplerian eccentric model. \emph{Left panel}: RV versus phase for \planetb\ ($5.3518$\,days). Blue markers represent HARPS-N measurements, orange and green markers represent ESPRESSO 1 and 2 measurements, respectively, and magenta squares are the phase-binned data. The black curve represents the best model. White noise has been added to the errors. \emph{Middle panel}: Same as the left panel, but for the second Keplerian at $15.678$\,d. \emph{Right panel}: transit of \planetb\ with blue points the \emph{K2} data, magenta squares the phase-binned data, and the black line its best model. Bottom panels represent the residuals of each corresponding model.}
    \label{fig:orbit}
\end{figure*}

We performed a global fit to the data (the \emph{K2} photometry and HARPS-N and ESPRESSO RVs) using the \texttt{PyORBIT} code \citep{Malavolta16}. We used \texttt{MultiNest} for the parameter inference with 1000 live points. The photometric transit was modeled using the \texttt{batman} transit model \citep{Kreidberg15}, quadratic limb darkening, and an exposure time of $1764.944$\,s for the photometry to account for the long-cadence observations following \citet{Malavolta18}. Eccentricity $e$ and longitude of periastron $\omega$ were combined in the fitting parameters $\sqrt{e}\cos\omega$ and $\sqrt{e}\sin\omega$ as per the recommendations of \citet{East13}. All priors were either uniform or log-uniform (the latter in the case of the semi-amplitudes for all Keplerians and the periods of the non-transiting Keplerian signals) and are listed in Table \ref{tab:globalfit}. White noise, additional to the errors, and an offset was included per instrument, where the ESPRESSO 1 and ESPRESSO 2 data were considered to be from two different instruments.

Given the strong periodicity of $15.68$\,d seen in the RV BGLS periodogram, we ran a model including two Keplerians, one relating to the transiting planet \planetb, and one relating to this periodicity of 15.678\,d (the period was constrained to be between 10 and 20 days to speed up the \texttt{MultiNest} run). We allowed both orbits to be eccentric. The resulting orbits are shown in Figure \ref{fig:orbit}, the parameters listed in Table \ref{tab:globalfit}, and the corner plots of the fit parameters shown in Figures S3 and S4 in the online supplementary material.

\begin{table}
\centering
\caption{\thisstar\ system parameters from combined fit.}            
\label{tab:globalfit}
\begin{tabular}{l l l } 
\hline\hline                 
Parameter & Prior & Value\\       
\hline
\multicolumn{3}{l}{\emph{Stellar parameters}}\\
\hline
$\rho_{*}$ [$\rho_\odot$] & $\mathcal{N}(0.43,0.02)$ \\
LD coefficient $q_{1}$  & $\mathcal{U}(0,1)$ &  $0.55\pm0.28$ \\
LD coefficient $q_{2}$  & $\mathcal{U}(0,1)$ &  $0.42_{-0.28}^{+0.33} $  \\
Systemic velocities: \\
$\gamma_{HN}$ [m/s] & $\mathcal{U}(-16280,-16270)$ &  $ -16275.2\pm0.3 $ \\
$\gamma_{E1}$ [m/s] & $\mathcal{U}(-16450,-16350)$ &  $ -16405.3\pm0.6 $ \\
$\gamma_{E2}$ [m/s] & $\mathcal{U}(-16450,-16350)$ &  $ -16408.5\pm0.4 $ \\
White noise: \\
$s_{\rm j,HN}$ [m/s] & $\mathcal{U}(0.01,10)$ &  $1.50_{-0.46}^{+0.42}$ \\ 
$s_{\rm j,E1}$ [m/s] & $\mathcal{U}(0.01,10)$ &  $1.75_{-0.67}^{+0.77}$ \\ 
$s_{\rm j,E2}$ [m/s] & $\mathcal{U}(0.01,10)$ &  $1.39_{-0.32}^{+0.38}$ \\ 
$s_{\rm j,K2}$ [ppm] & $\mathcal{U}(1,6100)$ &  $7_{-4}^{+5}$ \\ 

& \\
\hline
\multicolumn{3}{l}{\emph{Transit and orbital parameters} } \\
\hline
$P_b$ [d] & $\mathcal{U}(5.34,5.36)$ &  $5.3518\pm 0.0004 $ \\
$T_{ \rm tr, b}$ [BJD -2450000] & $\mathcal{U}(7100.0,7100.2)$ &  $7100.0768_{-0.0018}^{+0.0019} $ \\
$T_{\rm 14}$ [d] &  &  $0.133_{-0.020}^{+0.015} $ \\
$R_{\rm p,b}/R_{*}$ & $\mathcal{U}(0.00001,0.5)$ &  $0.01346_{-0.00061}^{+0.00074} $ \\
$i_b$ [deg] &  &  $86.43_{-0.21}^{+0.37} $ \\
Impact parameter $b_b$ & $\mathcal{U}(0,1)$ &  $0.66_{-0.12}^{+0.11} $ \\
$\sqrt{e_b}~\cos{\omega_b}$ & $\mathcal{U}(-1,1)$ &  $0.11_{-0.14}^{+0.15} $ \\
$\sqrt{e_b}~\sin{\omega_b}$ & $\mathcal{U}(-1,1)$ &  $-0.25_{-0.23}^{+0.33} $ \\
$e_b$ &  &  $0.13_{-0.09}^{+0.13} $ \\
$\omega_b$ [rad] &  &  $-1.21_{-0.48}^{+1.29} $ \\
$K_b$ [m/s] & $\mathcal{LU}(0.01,100)$ &  $2.21\pm0.32 $ \\
$P_c$ [d] & $\mathcal{LU}(10,20)$ &  $15.6785_{-0.0063}^{+0.0064} $ \\
Phase $\phi_{c}$ [rad] & $\mathcal{U}(0,2\pi)$ &  $1.38 \pm 0.20 $ \\
$\sqrt{e_c}~\cos{\omega_c}$ & $\mathcal{U}(-1,1)$ &  $0.05_{-0.22}^{+0.21} $ \\
$\sqrt{e_c}~\sin{\omega_c}$ & $\mathcal{U}(-1,1)$ &  $0.00\pm0.20 $ \\
$e_c$ &  &  $0.07_{-0.05}^{+0.07} $ \\
$\omega_c$ [rad] &  &  $-1.39_{-2.62}^{+1.68} $ \\
$K_c$ [m/s] & $\mathcal{LU}(0.01,100)$ &  $3.27_{-0.32}^{+0.31} $ \\
& \\
\hline
\multicolumn{3}{l}{\emph{Planetary parameters}} \\
\hline
$M_{\rm p,b} ~[\rm M_\oplus]$  &  &  $5.29_{-0.77}^{+0.76} $ \\
$R_{\rm p,b} ~[\rm R_\oplus]$  &  &  $1.82_{-0.09}^{+0.11} $ \\
$\rho_{\rm p,b}$ [$\rm g\;cm^{-3}$] &  &  $4.8\pm1.0 $ \\
$a_b$ [AU] &  &  $0.0570\pm0.0012 $ \\
$M_{\rm p,c}\sin i ~[\rm M_\oplus]$  &  & $11.3\pm1.1 $ \\
$a_c$ [AU] &  &  $0.1166\pm0.0025 $ \\
\hline       
\hline
\vspace{-0.3cm}
\end{tabular}
\newline \centering 
{\footnotesize Note: For the priors, $\mathcal{N}$ indicates a Gaussian prior with mean and standard deviation. $\mathcal{U}$ indicates a uniform prior and $\mathcal{LU}$ a log-uniform prior with logarithm base 2 where the values in brackets indicate the minimum and maximum value.}
\end{table}

\begin{figure}
    \includegraphics[width=\columnwidth]{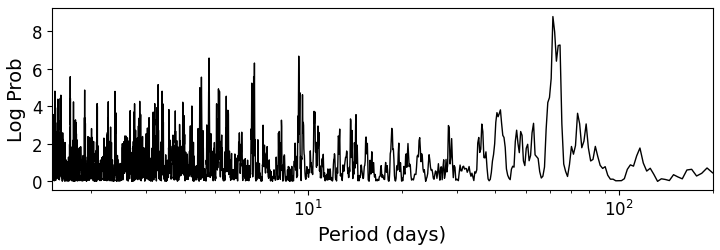}
    \caption{BGLS periodogram of the residual RVs after subtracting a 2-Keplerian eccentric model.}
    \label{fig:residuals}
\end{figure}

The fit converges to two low-eccentricity orbits, with the transiting planet ($P=5.3518$\,d) fitted with a slightly higher eccentricity ($e=0.13$) than the second signal at $15.678$\,d ($e=0.07$). Both eccentricities are, however, consistent with zero at the 2-sigma level. The transiting planet \planetb\ has its radius ($1.82$\,\rearth) determined with 20 sigma significance. The mass of \planetb\ is fitted as $5.29$\,\mearth, significant to almost 7 sigma. 

The best-fit white noise was between $1.4$ and $1.8$\,m/s depending on the instrument, with the first season of ESPRESSO having the largest white noise. This value is consistent with the estimates derived in Section \ref{activity}. We ran the same model multiple times to check whether the nested sampling algorithm found consistent results, which was the case. The residual RMS of the HARPS-N and ESPRESSO data, respectively, is $3.3$ and $2.1$\,m/s. This is in agreement with the expected RMS given the RV errors and fitted white noise.

A fit with only the transiting planet was run in order to assess the model evidences. A 2-Keplerian model was very strongly favoured with Bayes factors exceeding $10^{10}$. Similarly, a model was run where the two Keplerians were kept circular. This circular model was preferred but only with a Bayes factor of 51, which is not decisive (\citet{Kass95} recommend a Bayes factor higher than 150). Since there is no physical reason to prefer circular orbits over moderately eccentric ones for \thisstar, we chose to stick with the eccentric solutions, even though both fitted eccentricities are fully consistent with zero.

In the next sections, we discuss the mass accuracy of \planetb, as well as the origin of the strong sinusoidal signal at $15.678$\,d. The residuals of this two-Keplerian model also show some periodicity between 60 and 65 days and lots of forested peaks of periodicity at periods lower than 10 days, as shown from the BGLS periodogram in Figure \ref{fig:residuals}. We discuss this further in Section \ref{extra}.

\section{Accuracy of the mass of K2-111 b}\label{accuracy}

The mass of the transiting planet \planetb\ is precisely determined, with 1 sigma errors one seventh of the median value. However, it is important to also assess the accuracy of the fitted mass. Inadequate sampling and/or models can influence the accuracy of the extracted parameters and in particular the semi-amplitudes and thus planetary masses. An extreme example of this is Kepler-10\,c, where HIRES and HARPS-N data found very discrepant mass estimates \citep{Dum14,Weiss16}. Investigating this, \citet{Raj17} showed that the discrepancy could possibly be explained by sub-optimal sampling and/or the model choice.

We analysed the HARPS-N and ESPRESSO data separately to assess the mass accuracy of \planetb. The larger number of data points from HARPS-N could be balanced against the higher precision of the ESPRESSO data making both data sets qualitatively comparable. It is thus reasonable to expect comparable results by analysing the data sets separately. We ran a global fit (including the \emph{K2} photometry) assuming a 2 Keplerian eccentric model on the separate data sets. Similar priors, boundaries, and live points were used as for the full global fit. Using the HARPS-N data, we found a planet mass $M_{p,b} = 4.40^{+1.05}_{-1.03}$ Earth masses, while the ESPRESSO data converged to $M_{p,b} = 6.85^{+1.29}_{-1.17}$ Earth masses. These are different at the $1.5$ sigma level with, unsurprisingly, the mass as constrained by the full data set in between the values. We thus proceeded to check whether we could explain that difference.

The most obvious check would be to investigate the effect of changing the model. First, we looked at the inclusion of a third Keplerian in the model, given the periodicities present in the residuals of the global 2-Keplerian model. We ran additional models assuming three Keplerian signals, as explained in more detail in Section \ref{extra}. The mass determinations of \planetb\ were well within one sigma, with $4.52$ and $6.89$ Earth masses from the HARPS-N and ESPRESSO data, respectively, and $5.67$ Earth masses from the full data set. We have also run multiple models including a GP regression with a quasi-periodic kernel, as explained in detail in Section \ref{origin}. The mass of \planetb\ stays well within one sigma of the 2-Keplerian model in all the cases, regardless of the number of included Keplerian signals additional to the GP. We conclude that changing the model (within reason) does not explain the modest discrepancy for the planet mass of \planetb.

Next, we looked at the sampling. Given the multiple periodicities present in the data, the near-3:1 resonance of these periods, and the unknown effects from stellar activity, sampling could be problematic. Both data sets have been sampled fairly differently. Within one season, the HARPS-N data has a median spacing of 1 day between measurements, with multiple dense series of observations. The ESPRESSO data has a median spacing of 5 days between measurements with individual observations being much more spaced out. 

Inspired by the simulations from \citet{Raj17}, we created synthetic data and used our real observing calendars and errors to extract the semi-amplitudes of the inserted signals. We have done so in an analytic manner to speed up computational time by keeping the periods fixed and analytically solving for the minimum chi-squared, weighted by the errors. 

To create the synthetic data, we assumed circular orbits and used the periods and semi-amplitudes from the global fit of the full data with two circular Keplerians ($P_1=5.351962$\,d, $K_1=2.17$\,m/s and $P_2=15.679213$\,d, $K_2=3.23$\,m/s). We sampled uniformly over 50 different phases between $-\pi$ and $\pi$ for both signals. The analytical solution was obtained assuming $1$\,m/s additional white noise, added in quadrature to the errors\footnote{Not adding white noise did not affect the outcome.}. No noise was added to the individual data points meaning that if the data are sampled well, the semi-amplitudes should be extracted exactly as inserted. We used 4 different time series: the ESPRESSO observing calendar, the HARPS-N observing calendar, the combined calendar, and a uniform calendar using 154 data points sampled uniformly over the timespan of the full data set and with uniform errors. We find that all observing calendars result in well-extracted semi-amplitudes, within 1\,cm/s for both signals. The very different sampling of our data sets should thus not necessarily matter either.

As in \citet{Cloutier19}, we tried to identify any potentially anomalous points. For the fixed periods $P_1 = 5.3518$\,d and $P_2=15.678$\,d (corresponding to the two strongest signals in our data), we computed the BGLS periodogram probability and the associated fitted semi-amplitude and uncertainty \citep[see ][ for details on the computation]{ME15,ME17}. We ran this 154 times, always leaving one data point out. We can thus compare how each data point affects the period probability and corresponding fit. From all data points, there is one data point that stands out. Leaving out that data point results in a fitted semi-amplitude that is different than all the other fitted semi-amplitudes (it is lower for both periods). We do note that, while an obvious outlier, the difference is just within the fitted semi-amplitude errors. The data point is from the second set of ESPRESSO data at JDB$=2458821.675673$\,d.

We ran extra models excluding this data point, both for the ESPRESSO data set and the full data set. As expected, the planet mass of \planetb\ goes down, but only marginally (from $5.29$ to $5.11$ Earth masses for the full data set). The one slightly anomalous ESPRESSO data point can thus not explain the difference either.

Given the different time span of the HARPS-N and ESPRESSO RV data, the small fitted mass difference could potentially arise from librating co-orbital bodies. We thus performed a co-orbital analysis for \planetb. We followed the theoretical framework described in \citet{Leleu17} and subsequently applied in other observational works \citep[e.g. ][]{Lillo-box18a,Lillo-box18b,Toledo20} to constrain the presence of planet-mass co-orbital bodies. This framework introduces the parameter $\alpha$ in the usual RV equation which corresponds to the mass ratio between the trojan and the planet. If compatible with zero, the data are unable to constrain the presence of co-orbital bodies. If significantly different from zero and positive (negative) the signal contains hints for the presence of a mass imbalance at the L$_5$ (L$_4$) Lagrangian point (located at $\pm 60^{\circ}$ from the planet location and on the same orbital path in the restricted three-body problem). We assumed wide Gaussian priors on the planet parameters (using the two-Keplerian circular model for simplicity). We find that $\alpha = 0.27^{+0.25}_{-0.21}$, which is compatible with zero at the $95\%$ confidence level. The distribution is, however, $1.3\sigma$ shifted towards positive values, indicating a possible mass imbalance at the L$_5$ location. The mass of a trojan located exactly at this Lagrangian point and causing this imbalance would be $1.7^{+1.6}_{-1.3}$\,\mearth, which allows us to discard at the $95\%$ confidence trojans more massive than $4.5$\,\mearth\ at L$_5$ and $0.4$\,\mearth\ at L$_4$. Assuming coplanarity with the planet orbit, the median mass for this trojan would imply a transit depth of $\sim70$ ppm, at the limit of the photometric sensitivity from the \emph{K2} data. An inspection of the phase-folded light curve does not show any dimming of this depth  at the L$_5$ location or at any other location, although the photometric sensitivity, the possibility of libration and mutual inclination with the planet's orbital plane leave the possibility of the co-orbital case still open.

Finally, we considered that the difference could be explained by unaccounted for effects of stellar variability. The WASP photometry and the ESPRESSO S-index both showed signs of periodicity in the vicinity of the transiting planet period. We have tried training a GP with a quasiperiodic kernel to the S-index data, but the posterior distributions were essentially always equal to the uniform prior distributions, even if we narrowed the 'rotation period' hyperparameter to be close to the peak seen in the BGLS periodogram. We found no obvious way of modelling possible stellar activity effects around the $5.35$\,d period, but hypothesise it may be the explanation for the differences seen from both data sets.

From these tests we conclude that the planet mass of \planetb\ is as accurate as we can confidently infer from the available data.

\section{The origin of the 15.678 day signal}\label{origin}

For periodic signals in RV data, unrelated with known transiting planets, it is important to understand the origin of this signal. This can be mainly due to two things: a non-transiting planet or a signal created by stellar activity on the surface of the star \citep[see e.g. ][for a good discussion on this]{Faria20}.

Signals arising from stellar activity are often mimicked in activity indicators, such as the FWHM of the CCF or the S-index, even for moderate, or low-activity stars. A correlation coefficient between RVs and indicators may be lowered due to time shifts between the time series \citep[see e.g.][]{San14,ACC19}, but similar periodicities are often seen in the data. For example, high-resolution observations from the Sun-as-a-star clearly show similar periodicity structure in the RVs and the indicators \citep{ACC19,Maldonado19} around the rotation period and its harmonics while the correlation coefficients are lowered due to a 2-day time shift. It is therefore important that the $15.678$\,d signal appears so strongly in the RVs but is not seen in the BGLS periodograms of any indicator (Figure \ref{fig:BGLS}). Overall the indicators do not show strong periodicity at all while the RV signal in both the HARPS-N and ESPRESSO data is strong, consistent and stable over time. This points to the $15.678$d signal being planetary.

Some photometry data we gathered does, in contrast with the spectral indicators, show more clear signs of periodicity around 30 days and its harmonic around 15 days. This could thus point to the signal seen in the RVs being stellar in origin. However, it is important to point out that, while there is some modest forested structure around 30 days in the HARPS-N RVs, the ESPRESSO RVs show no sign of a 30d periodicity. If the $15.678$\,d signal is related to stellar activity, it would be surprising \citep[but perhaps not impossible - see ][]{Nava20} to see strong periodicity at the harmonic and none around the rotation period. Additionally, the \emph{K2} data, while shorter in time span, firmly rules out variations greater than 500\,ppm at this period.

Assuming the $15.678$\,d signal may be stellar in origin, we used \texttt{PyORBIT} to model the RV data with a model including just the transiting planet, a GP with a quasi-periodic kernel (similar to what we did for the photometry in Section \ref{activity}) and additional white noise. To speed up the computation, we modelled only the RV data and put priors on the parameters of the transiting planet based on a fit of the \emph{K2} photometry only.

As a first run, we put additional priors on the rotation period, decay timescale and coherence scale based on the \texttt{PyORBIT} results of modelling all the photometry (see Section \ref{activity}). The model converged to a reasonable solution. We compared the evidences of this model with a model of two Keplerians and white noise (also with a transiting planet prior and not using \emph{K2} data). The two Keplerian model is decisively preferred with a Bayes factor exceeding $10^6$.

We also ran \texttt{PyORBIT} while putting no priors on the kernel hyperparameters and allowing the rotation period to be below 100\,d. The two Keplerian model is still preferred over this model, but the Bayes factor is only 16 now. The hyperparameters were constrained to $P_{\text{rot}}= 62.67^{+0.07}_{-46.99}$\,d, $P_{\text{dec}} =1594^{+289}_{-451}$\,d, $h=3.55^{+1.61}_{-0.85}$\,m/s, and $w=0.21^{+0.84}_{-0.05}$. It stands out that the decay timescale goes to very high values, comparable with the timespan of the RV data (which is 1590 days). The posterior distributions of the rotation period and coherence scale are both multimodal with peaks close to 60 and 15 days and corresponding coherence scales of $0.20$ and 1, respectively. We thus ran three additional models where the rotation period hyperparameter was constrained to be around 15, 30 or 60 days. We saw the same result across all these models. The model evidences were lower than the two Keplerian model (with Bayes factors ranging from 4 to 247) and the decay timescales converged to be higher than 1000 days. 

It could of course still be the case that the periodicity between 15 and 16 days is a mixture of both stellar and planetary origin. We thus also ran a 2-Keplerian model with an additional quasiperiodic GP. Putting priors on the kernel hyperparameters from the photometry resulted in a good fit and an amplitude kernel hyperparameter of $h=1.28\pm0.33$\,m/s. The 2-Keplerian model without including a GP is still mildly preferred with a Bayes factor of 10. Importantly, the recovered mass for the $15.678$\,d Keplerian signal, if it were due to a planet, remains the same, within 1 sigma of all the other models we tried. 

\begin{figure}
    \includegraphics[width=\columnwidth]{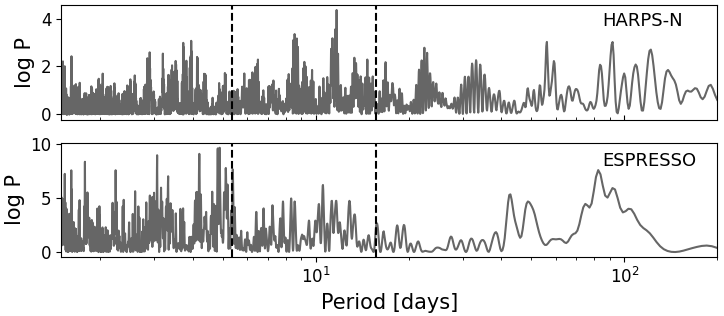}
    \caption{BGLS periodogram of the difference between the red and blue RVs - Log Probability is plotted against period. Top plot is for HARPS-N data, bottom plot for ESPRESSO data. The vertical dashed lines represent the two strongest periodicities in the RVs.}
    \label{fig:CRV}
\end{figure}

Finally, we checked the coloured RVs as extracted from HARPS-N and ESPRESSO. Effects from stellar variability are not stable across wavelength while a signal arising from a planet should be the same regardless of the wavelength \citep[e.g. ][]{Figueira10,Hue08,Reiners10,Zechmeister18}. The red, green, and blue RVs from ESPRESSO show the same overall structure in the BGLS periodogram with peaks around the transiting planet period and around $15.678$\,days. For the HARPS-N coloured RVs, each set shows a different overall structure, but all three show strong probability around $15.678$\,d. We subtracted the red RVs from the blue RVs and plotted the BGLS periodogram of these residuals in Figure \ref{fig:CRV} for each data set. Signals arising from planets should be fully removed in these residuals while stellar activity effects will remain. The probability for the $15.678$\,d signal is within the noise for both the HARPS-N and the ESPRESSO data making the planet hypothesis more likely than the activity hypothesis.

Interestingly, the strongest periodicities in the ESPRESSO coloured RV residuals are around 5\, days and below. There may thus indeed be some excess signal in the transiting planet RV fit from the ESPRESSO data that comes from these short-timescale effects, as hypothesised in Section \ref{accuracy}.

Given the highest, albeit sometimes marginal, evidence for the two Keplerian model, the high decay timescale in the model with 1-Keplerian and a GP, the mass consistency for all 2-Keplerian models, and the consistency across wavelength, we conclude that the $15.678$\,d signal is stable, at least over the timescale of the data and it points towards the signal being planetary in origin rather than stellar. We use \planetc\ for this candidate planet for the remainder of this work. 

\begin{figure*}
    \includegraphics[width=\textwidth]{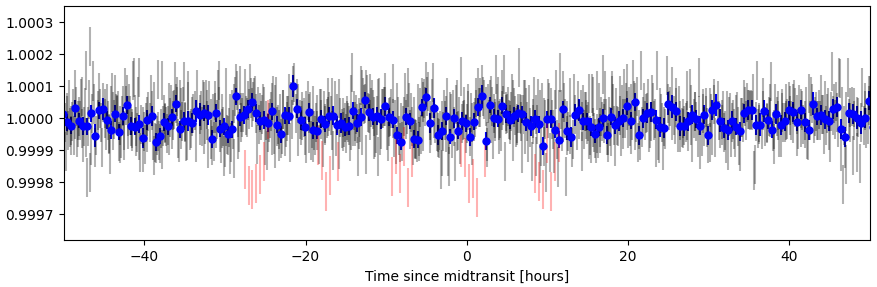}
    \caption{Flux versus time of the \emph{K2} light curve, phase-folded with the orbital period and phase of \planetc. Blue dots are binned data, binned over 30 minutes. Grey and red errors are shown from the unbinned data where the red errors are the photometric points inside the transit of \planetb. The expected transit depth of \planetc, assuming a mass-radius relation from \citet{Chen17}, is $0.0006$, well below the lowest point in the figure.}
    \label{fig:transitc}
\end{figure*}

We checked the \emph{K2} light curve for signs of \planetc. The light curve, phase-folded with the orbital period and phase mentioned in Table \ref{tab:globalfit}, is shown in Figure \ref{fig:transitc}. Assuming the probabilistic mass-radius relation of \citet{Chen17}, \planetc\ is expected to have a radius of $3.37$\,\rearth. This would result in a transit depth of $0.0006$. If this planet would transit we would be able to easily detect it in the \emph{K2} light curve. With no points lower than $0.9997$ in the light curve, it is clear that \planetc\ does not transit. Assuming co-planarity between the two planets, the impact parameter of \planetc\ would be $1.25$, indicating we would not expect this second planet to transit. 

Both planets orbit \thisstar\ in near 3:1 mean-motion resonance. This could induce transit timing variations (TTVs) for the transiting planet \planetb. We checked for this by fitting for each transit separately. Since the star was observed in long cadence, the amount of data points in each individual transit is low. We thus kept the depth of each transit fixed to the value obtained from the global fit and only allowed the central time of transit to vary. All transits occurred on the predicted time from a constant period, fully consistent with a null result on TTV, however, as expected, errors on the TTVs are quite large, on the order of 10 minutes. We used \texttt{REBOUND}\footnote{\url{http://github.com/hannorein/rebound}} \citep{Rein12}, an N-body integrator, to estimate the expected TTV variations given our orbital solution. We found that the maximum expected amplitude of TTV variations would be 2 minutes, well below our precision and fully consistent with our values being close to zero.

\section{Additional signals in the data}\label{extra}

As mentioned in Section \ref{global}, there is some periodicity left around 60 days in the residuals of the global 2-Keplerian model. We thus fitted the RV data and \emph{K2} photometry with a 3-Keplerian model which converged to a good solution. The 3-Keplerian model is slightly preferred, but only with a Bayes factor of 46, not enough to decisively say the 3-Keplerian model is more likely. From Figure \ref{fig:CRV}, we can see that neither data set shows strong signs of chromaticity for periods between 60 and 70 days. If the third Keplerian were due to a planet, it would have a minimum mass of 7 Earth masses. The masses of \planetb\ and \planetc\ do not change within 1 sigma of the 2-Keplerian solution.

When looking more closely at the solution of this 3-Keplerian model, the posterior distribution for the period of the third Keplerian is bimodal with a solution around $61.5$\,d and one around $64.5$\,d. These two period are likely aliases from each other, due to the time span of the data. Both solutions have similar probabilities and similar orbital parameters and it is unclear how to favour one over the other. Judging solely from phase coverage, we can say that the Keplerian solution around $61.5$\,d is less well sampled in phase space than the $64.5$\,d solution. 

Interestingly, the BGLS periodogram of the residuals from the 2-Keplerian model including a GP shows less probability around 60 days. We thus also ran a 3-Keplerian model including a GP. The model evidence was only a bit lower than the 2-Keplerian model. Overall, the orbital parameters of all three Keplerians were similar than for the model without the GP. The posterior distribution of the period is, however, no longer bimodal and only the $61.5$\,d, less-well sampled, solution is found.

It would not be surprising if there were more planets in this system, in addition to \planetb\ and \planetc. The current data however does not allow us to draw strong conclusions about the existence of these additional planets.

\section{Discussion and conclusion}\label{discussion}

\begin{figure}
    \includegraphics[width=\columnwidth]{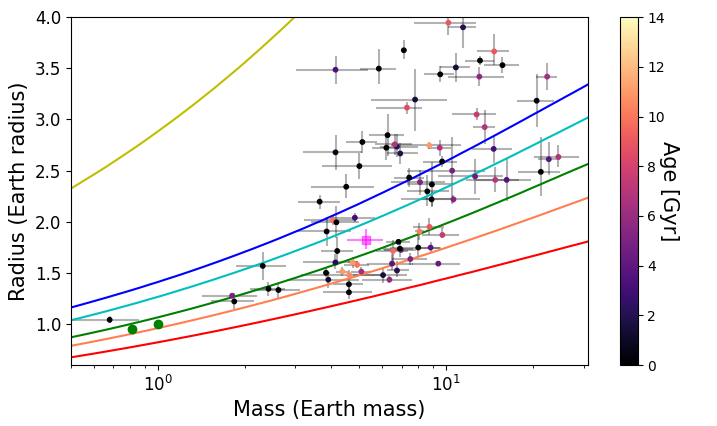}
    \caption{Mass-radius diagram of all planets smaller than $4$\,\rearth\ with a mass measurement significance better than 3 sigma (using data from the NASA exoplanet archive accessed 11 June 2020). The points are colour-coded according to their stellar host ages (in case no age was determined, the point is black). The green dots bottom left represent Venus and Earth, respectively. The solid lines show planetary interior models for different compositions, top to bottom: Cold H$_2$/He, 100\% H$_2$O, 50\% H$_2$O, 100\% MgSiO$_3$, 50\% Fe, 100\% Fe. The larger magenta square represents \planetb.}
    \label{fig:MR}
\end{figure}

In this work, we have confirmed the presence of the transiting planet \planetb. We have refined the planetary parameters, especially the planetary mass. The planetary radius, $1.82^{+0.11}_{-0.09}$ Earth radii, is significant to 20 sigma. Our fitted planet radius is marginally lower than the one reported in \citet{Fridlund17} ($R_p = 1.9\pm0.2$\,\rearth) but twice as precise. This is almost solely due to the more precise and accurate stellar radius, possible with the Gaia DR2 parallax measurement. We fitted the planet mass of \planetb\ to $5.29^{+0.76}_{-0.77}$ Earth masses, significant to almost 7 sigma. Our fitted value is about 3 Earth masses lower than the mass reported by \citet{Fridlund17}, but we note that they had only measured the planet mass to within 2 sigma ($M_p = 8.6\pm3.9$\,\mearth).

We also announce the presence of an additional non-transiting planet in this system, \planetc, at $15.678$ days with a minimum mass of $11.3\pm1.1$ Earth masses. While not confirmed with an independent technique, we have shown in Section \ref{origin} that the detected RV signal is unlikely to have arisen from stellar surface phenomena.

We have used two RV data sets that are qualitatively equal, with more data points from HARPS-N and higher precision from ESPRESSO. These data sets independently detected the two planets, with mass measurements significant to at least 4 sigma, though discrepant at the $1.5$ sigma level. Combining these two exquisite data sets made the detections much more significant and strengthened the planet hypothesis for the $15.678$\,d signal, now called \planetc.

\citet{Fridlund17} had hypothesised the presence of a more massive object with a period longer than 120 days in their work, from a fitted linear trend in their model. Neither the HARPS-N nor the ESPRESSO data show any signs of a long-term trend or indicate the presence of such a massive outer companion. We think the appearance of a linear trend in the data used by \citet{Fridlund17} was the result of the short data time span and very sparse sampling for this resonant planetary system. The period of \planetc\ is a bit less than three times the period of the transiting planet making these two planets orbit in near-3:1 resonance. 

We do see signs of a potential additional super-Earth with a period around $61.5$ days and a minimum mass of $7$ Earth masses. The current data set is not sufficient to draw conclusions about the nature of this signal. It is, however, interesting to note that, if this is indeed a planet, the period would be about 4 times the period of \planetc, making the three planets fit nicely in a near-resonant chain.

Given the near-3:1 resonance of the two planets in this system, we investigated the dynamical stability. As also illustrated in \citet{Fridlund17}, for the estimated mass of \planetc, we'd expect the system to be stable unless the eccentricity exceeded $e \sim 0.3$ \citep{Petrovich15}. To check this, we ran a small suite of N-body simulations using \texttt{MERCURY6.2} \citep{Chambers99} with the Bulirsch-Stoer integrator. We set the orbital parameters and planet masses, by randomly sampling from the ranges presented in Table \ref{tab:globalfit}. These simulations indicate that the planetary system is indeed stable on long timescales (well beyond $10^8$\,yr).

We also used these simulations to check if the periodicity at around 60 days in the residuals of the global 2-Keplerian model could be due to some kind of resonant interaction between the 2 inner planets. From 5 of the N-body simulations, we considered a period of 1000 days and randomly extracted $\sim 200$ stellar RVs. We then added noise to these RVs with a magnitude similar to that of the jitter presented in Table \ref{tab:globalfit}, and assumed that the uncertainties were similar to those presented in Table \ref{table:rvshort}.

We first used \texttt{PyORBIT} to check that it would recover the 2 known planets in this synthetic data. This was the case, with a similar precision to that presented in Table \ref{tab:globalfit}. We then used \texttt{PyORBIT} to check if it could recover a statistically significant 3 planet solution, assuming that the third planet has a period $P > 30$ days. All of these analyses recovered the 2 known planets, but none converged on a 3 planet solution. Therefore, it seems unlikely that the long-period signal is due to a resonant interaction between the 2 inner planets.

The \thisstar\ system is also intriguing as the host star, with an age at least older than 10\,Gyr, is among the oldest stars in the Universe. Given this old age, it is possible that the planetary orbits, and especially the inner one, would have circularised \citep[see e.g. ][]{Goldreich66,Jackson08}. This is in line with the eccentricities being fully consistent with zero. The orbits are in near mean-motion resonance which could pump the eccentricities of the orbits as well, impeding tidal circularisation \citep[see e.g.][]{Beust96}. Indeed there are other examples of close-in planets on eccentric orbits around very old stars \citep[e.g. ][]{Motalebi15,ME16}.

Not much is known about planetary systems around the very oldest stars and the history of their formation. It is therefore crucial that we study these systems, as also pointed out by \citet{Fridlund17}. We checked the NASA Exoplanet Archive\footnote{\url{https://exoplanetarchive.ipac.caltech.edu} - accessed on 11th June 2020}. Roughly half of the >4000 planets listed in the archive mention a stellar age for the planet host. Only 57 of these, orbiting 39 different stars, have a stellar host age greater than 10 Gyr, including \thisstar. We do caution that most stellar ages are badly constrained. When we take into account that our host star is also iron-poor and select only those systems with sub-solar overall metallicity, there are only 22 planets left, orbiting 15 stars, with a wide variety of orbital and planetary parameters. 

Figure \ref{fig:MR} shows the mass-radius diagram for \planetb\ together with all small planets ($R_p < 4$\,\rearth) where a mass measurement is known with a significance better than 3 sigma. \planetb\ sits in a region between the 50\% H$_2$O and 100\% MgSiO$_3$ composition lines \citep{Zeng13,Zeng16}, with a bulk density slightly lower than the Earth. There are no obvious trends noticeable with stellar age, which could be due to low-number statistics.

\subsection{Interior structure of K2-111 b}

The host star, \thisstar, is iron-poor. This in itself is interesting as there are only a handful small planets around iron-poor stars that have their radius and mass both measured. More interestingly, however, is the significant alpha-enhancement of \thisstar. The ratios between iron, magnesium, and silicon are different to the solar ratios. As these elements are the core building blocks for a planet's core and mantle, the difference in these ratios is important as this could affect the interior structure of the small planets around \thisstar.

Using the stellar chemical composition we can infer the mass fraction of the planet building blocks in the \thisstar\ system, assuming the stellar composition is a good proxy for the disc composition at the time of planet formation. We can use our obtained chemical abundances from the spectra for this purpose. We followed the procedure described in \citet{San15,Santos17} which uses a simple stoichiometric model and chemical abundances of Fe, Mg, Si, C, and O to predict expected iron and silicates mass fractions of the planet building blocks. For \thisstar\ we find that the expected mass fractions of the iron and silicates building blocks are $21\%$ and $79\%$, respectively. The iron mass fraction is, as expected for this old, alpha-enhanced star, significantly lower than the iron mass fraction of the solar planet building blocks \citep{Santos17}.

With both a precise planetary mass and radius for \planetb, we can go a step further and perform a more detailed modeling of the stellar, planetary, and orbital characteristics to infer the possible interior structure of the transiting planet. The methodology we used is described in \citet{Alibert20} and is based on the models used by \citet{Dorn15,Dorn17}. The physical model has been improved since those works, e.g. by using a new equation of state for the water layer \citep{Haldemann20}. The gas envelope is treated using the model of \citet{LopezFortney14} which gives the thickness of the gas envelope as a function of the age, planetary mass, etc. We neglect in this analysis the effect of the gas envelope on the radius of the solid part. The analysis is now also done using neural networks exploring the probability distributions of the internal structure parameters. We refer to \citet{Alibert20} for more details.

The data going into the model are the stellar mass, radius, effective temperature, and age as well as the chemical abundances of Fe, Mg, Si and the planetary mass, radius, and period. The posterior distributions of the relevant internal structure parameters are shown in Figure S5 in the online supplementary material. We find that \planetb\ has a small gas envelope ($\sim0.01$\,\mearth\ and $0.1$\,\rearth\ of gas). This shows that neglecting the compression effect of the gas envelope onto the planetary interior, which the model assumes, is perfectly justified in our case. The planet has an iron core with a mass fraction of around 10\% (much smaller than the Earth), a big silicate mantle (up to more than 80\% of the planet mass could be in the mantle) while the water layer is less well constrained (the posterior distribution has a slight preference for a small mass fraction). This result is again in agreement with the star being iron-poor and is qualitatively similar to the results from the simple stoichiometry model. 

Due to the star being very old and the planet orbiting relatively close to the star, it is likely that an $0.1$\,\rearth\ H/He gas envelope around \planetb\ did not survive due to photo-evaporation. When we do not allow a gas envelope to be present in our model, this resulted in a very similar solution with a small iron core, a large silicate mantle and an unconstrained but likely small volatile layer. This is unsurprising given the gas envelope was found to be small and only $0.01$\,\mearth\ in the original model.

Finally, we explored whether the planetary and stellar parameters could be consistent with a truly terrestrial model where the planet interior structure is only composed of an iron core and a silicate mantle. We find that the stellar and planetary characteristics are consistent, within 2-sigma, with such a terrestrial 2-layer model for \planetb. The posterior distributions of these internal structure parameters are shown in Figure S6 in the online supplementary material. The iron core mass fraction in this scenario is 13\%, still significantly less than the Earth. The slightly higher mass as derived from the ESPRESSO data alone, could be favoured for this terrestrial two-layer model and make it more consistent than 2 sigma. If we allowed a gas envelope with the two-layer model, we found that the gas envelope was again fitted to very small values, ruling out a large gas envelope that could potentially be stable around a terrestrial planet.

We caution that these results are model-dependent and that more models with various physical ingredients should be explored, but \planetb\ could be an actual terrestrial planet, though with a much lower core mass fraction than the Earth, compatible with its host's stellar composition. The small core mass fraction is a feature likely shared by other planets around thick disc or halo stars, given their reduced iron mass fractions compared to planets around thin disc stars \citep[see e.g.][]{Santos17}. Recently, \citet{Michel20} showed that solid planets around thick disc or halo stars have a larger radii than planets around thin disc stars.

Planet formation and evolution theories improve by the discovery of more planetary systems around a variety of stars. To understand the small details of every aspect of planet formation and evolution, it is of the utmost importance we keep studying and characterising planetary systems in a wide parameter space, both for the stars and the planets. Precisely and accurately characterising systems like \thisstar, will refine our knowledge of planetary systems in the Galaxy.

\section*{Acknowledgements}

We thank the referee for a constructive report. We thank Josh Briegal and Jo\~ao Faria for helpful discussions.

AMo acknowledges support from the senior Kavli Institute Fellowships.
MRZO acknowledges financial support from the Spanish Ministry of Science and Innovation (MICINN) AYA2016-79425-C3-2-P.
JIGH acknowledges financial support from the Spanish MICINN under the 2013 Ram\'on y Cajal programme RYC-2013-14875. 
ASM acknowledges financial support from the Spanish MICINN under the 2019 Juan de la Cierva Programme. 
JIGH, ASM, RR, and CAP acknowledge financial support from the Spanish MICINN AYA2017-86389-P.
ACC acknowledges support from the Science and Technology Facilities Council (STFC) consolidated grant number ST/R000824/1.
CAW would like to acknowledge support from UK STFC grant ST/P000312/1.
PF was supported by MCTES through national funds (PIDDAC, PTDC/FIS-AST/32113/2017).
MP acknowledges financial support from the ASI-INAF agreement n. 2018-16-HH.0.
YA and JH acknowledge the Swiss National Science Foundation (SNSF) for supporting research through the SNSF grant  200020\_192038.
VA, EDM, SS, SCCB and ODSD acknowledge support from Funda\c{c}\~ao para a Ci\^encia e a Tecnologia (FCT)) through Investigador FCT contracts IF/00650/2015/CP1273/CT0001, IF/00849/2015/CP1273/CT0003, IF/00028/2014/CP1215/CT0002, IF/01312/2014/CP1215/CT0004 and DL 57/2016/CP1364/CT0004. 
This work was supported by FCT through national funds and by FEDER through COMPETE2020 - Programa Operacional Competitividade e Internacionaliza\c c\~ao by these grants: UID/FIS/04434/2019; UIDB/04434/2020; UIDP/04434/2020; PTDC/FIS-AST/32113/2017 \& POCI-01-0145-FEDER-032113; PTDC/FIS-AST/28953/2017 \& POCI-01-0145-FEDER-028953; PTDC/FIS-AST/28987/2017 \& POCI-01-0145-FEDER-028987.
Parts of this work have been supported by the National Aeronautics and Space Administration under grant No. NNX17AB59G issued through the Exoplanets Research Program.
Part of this work has been carried out within the framework of the NCCR PLanetS supported by the SNSF. 
The INAF authors acknowledge financial support of the Italian Ministry of Education, University, and Research with PRIN 201278X4FL and the "Progetti Premiali" funding scheme.
FAP and CL would like to acknowledge the SNSF for supporting research with ESPRESSO through the SNSF grants nr. 140649, 152721, 166227 and 184618 and with HARPS-N through the SNSF grants nr. 140649, 152721, 166227 and 184618. The ESPRESSO Instrument Project was partially funded through SNSF's FLARE Programme for large infrastructures. The HARPS-N Instrument Project was partially funded through the Swiss ESA-PRODEX Programme.
This publication was made possible through the support of a grant from the John Templeton Foundation. The opinions expressed in this publication are those of the authors and do not necessarily reflect the views of the John Templeton Foundation.
This project has received funding from the European Research Council (ERC) under the European Union's Horizon 2020 research and innovation programme (project {\sc Four Aces}; grant agreement No 724427).

This research has made use of the SIMBAD database, operated at CDS, Strasbourg, France, NASA's Astrophysics Data System and the NASA Exoplanet Archive, which is operated by the California Institute of Technology, under contract with the National Aeronautics and Space Administration under the Exoplanet Exploration Program.
Based on Guaranteed Time Observations collected at the European Southern Observatory under ESO programmes 1102.C-0744, 112.C-0958, and 1104.C-0350 by the ESPRESSO Consortium.
Based on observations made with the Italian Telescopio Nazionale Galileo (TNG) operated on the island of La Palma by the Fundacion Galileo Galilei of the INAF (Istituto Nazionale di Astrofisica) at the Spanish Observatorio del Roque de los Muchachos of the Instituto de Astrofisica de Canarias.
The HARPS-N project has been funded by the Prodex Program of the Swiss Space Office (SSO), the Harvard University Origins of Life Initiative (HUOLI), the Scottish Universities Physics Alliance (SUPA), the University of Geneva, the Smithsonian Astrophysical Observatory (SAO), and the Italian National Astrophysical Institute (INAF), the University of St Andrews, Queen's University Belfast, and the University of Edinburgh.
We acknowledge the University of Warwick for running the WASP Data Centre.
This paper includes data collected by the \emph{K2}\ mission. Funding for the \emph{K2}\ mission is provided by the NASA Science Mission directorate. Some of the data presented in this paper were obtained from the Mikulski Archive for Space Telescopes (MAST). STScI is operated by the Association of Universities for Research in Astronomy, Inc., under NASA contract NAS5-26555. Support for MAST for non-HST data is provided by the NASA Office of Space Science via grant NNX13AC07G and by other grants and contracts.
This work has made use of data from the European Space Agency (ESA) mission {\it Gaia} (\url{https://www.cosmos.esa.int/gaia}), processed by the {\it Gaia} Data Processing and Analysis Consortium (DPAC, \url{https://www.cosmos.esa.int/web/gaia/dpac/consortium}). Funding for the DPAC has been provided by national institutions, in particular the institutions participating in the {\it Gaia} Multilateral Agreement.

\section*{Data availability}
All underlying data is available either in the appendix/online supporting material or will be available via VizieR at CDS.



\bibliographystyle{mnras}
\bibliography{References}


\clearpage
\appendix
\onecolumn
\section*{\centering \Huge Supplementary Material}

This supplementary material to the paper 'K2-111: an old system with two planets in near-resonance' by Mortier et al. contains plots of the full time series of the RVs, indicators, and photometry, the corner plots for the best global fit and the interior structure model, the orbital solution for the circular fit, and the full RV and indicator data table. The photometry data as well as the RVs and activity indicators will be provided via Vizier at CDS.

\clearpage
\begin{figure}
    \includegraphics[width=\textwidth]{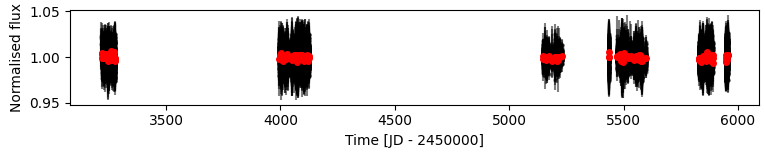}
    \includegraphics[width=\textwidth]{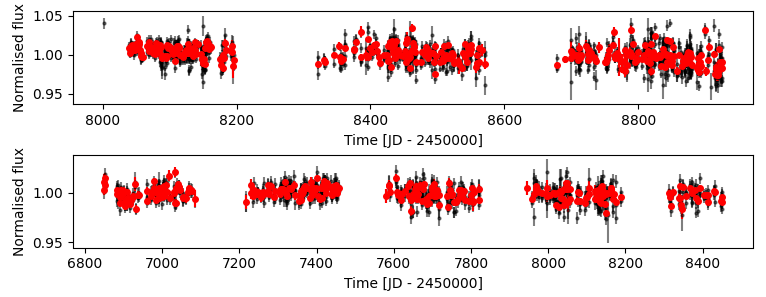}
    \caption{Full time series of the WASP and ASAS photometry time series. Grey points indicate all the data while the red points are daily binned measurements. {\emph Top to bottom} is the WASP, ASAS g and ASAS V data.}
    \label{fig:alldata}
\end{figure}

\begin{figure}
    \includegraphics[width=22cm,angle=90]{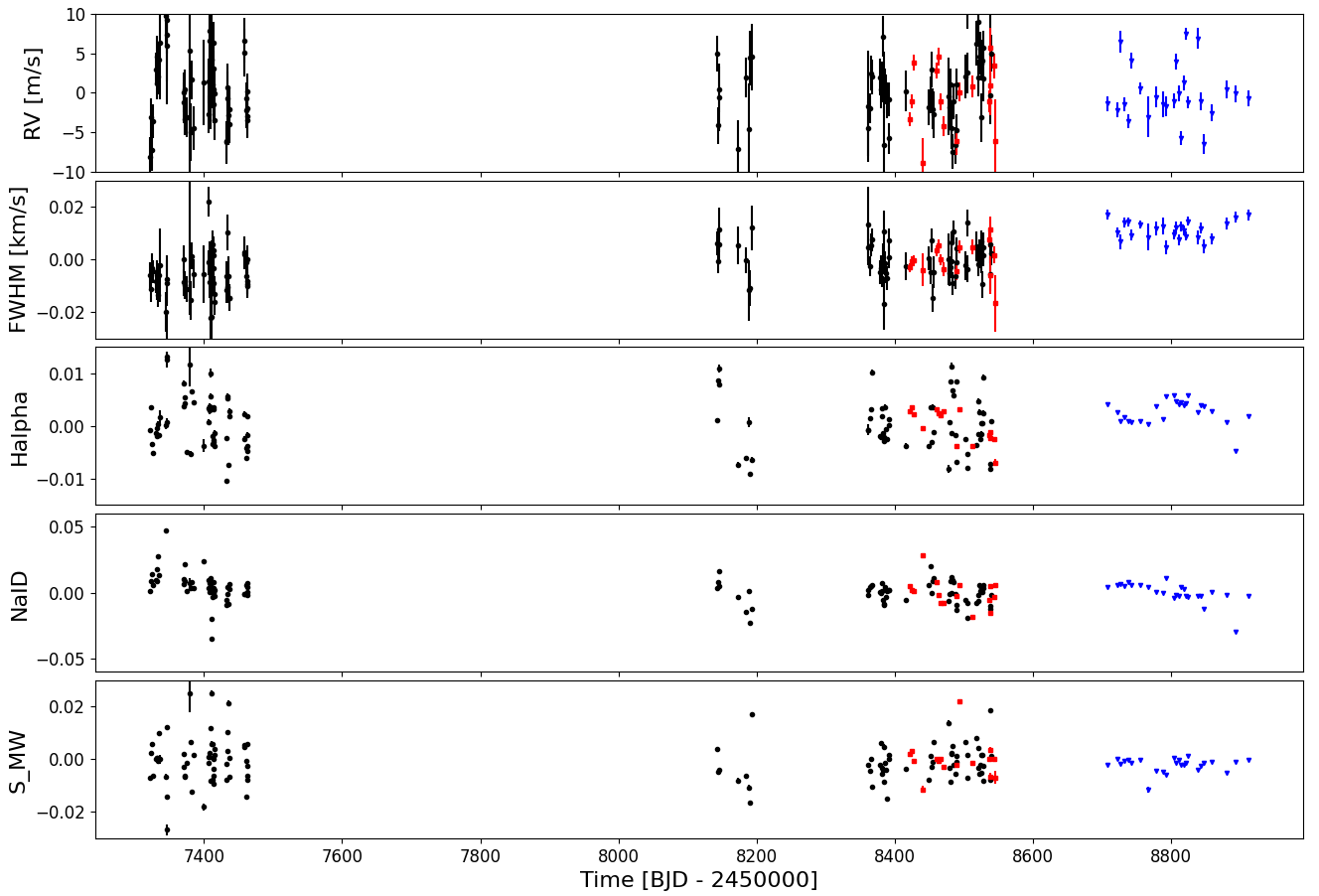}
    \caption{Full time series of the RVs and the indicators. All variations are relative around zero and the appropriate instrumental offsets have been subtracted for the RVS and FWHM. Black points indicate HARPS-N data, while red squares and blue triangles indicate ESPRESSO 1 and 2 data, respectively. Errors are shown for all data points, but are sometimes smaller than the marker size.}
    \label{fig:alldata}
\end{figure}

\begin{table}[h]
\centering
\caption{\thisstar\ system parameters from combined circular fit.}            
\label{tab:circularfit}
\begin{tabular}{l l }        
\hline\hline                 
\emph{Stellar parameters}  &  \\
\hline
Limb-darkening coefficient $q_{1}$  &  $0.54_{-0.28}^{+0.30}$ \\
Limb-darkening coefficient $q_{2}$  &  $0.42_{-0.27}^{+0.33} $  \\
Systemic velocities: \\
$\gamma_{HN}$ [m/s] & $ -16275.3\pm 0.3 $ \\
$\gamma_{E1}$ [m/s] & $ -16405.4\pm 0.6 $ \\
$\gamma_{E2}$ [m/s] & $ -16408.5\pm 0.4 $ \\
RV white noise: \\
$s_{\rm j,HN}$ [m/s] & $1.48_{-0.46}^{+0.41}$ \\ 
$s_{\rm j,E1}$ [m/s] & $1.77_{-0.65}^{+0.79}$ \\ 
$s_{\rm j,E2}$ [m/s] & $1.44_{-0.31}^{+0.37}$ \\ 
K2 white noise $s_{\rm j}$ [ppm] & $7 \pm 4$ \\ 

& \\
\hline
\emph{Transit and orbital parameters}  &  \\
\hline
$P_b$ [d] & $5.3520\pm 0.0003 $ \\
$T_{ \rm tr, b}$ [BJD -2450000] & $7100.0766\pm 0.0018 $ \\
$T_{\rm 14}$ [d] & $0.144_{-0.005}^{+0.006} $ \\
$R_{\rm p,b}/R_{*}$ & $0.01312_{-0.00044}^{+0.00041} $ \\
$i_b$ [deg] & $86.57_{-0.28}^{+0.35} $ \\
Impact parameter $b_b$ & $0.59_{-0.06}^{+0.04} $ \\
$e_b$ & $0.0 $ (fixed) \\
$\omega_b$ [rad] & $\pi/2 $ (fixed) \\
$K_b$ [m/s] & $2.17_{-0.32}^{+0.31} $ \\
$P_c$ [d] & $15.6792_{-0.0061}^{+0.0065} $ \\
Phase $\phi_{c}$ [rad] & $1.40 \pm 0.20 $ \\
$e_c$ & $0.0$ (fixed) \\
$\omega_c$ [rad] & $\pi/2$ (fixed) \\
$K_c$ [m/s] & $3.23_{-0.30}^{+0.33} $ \\
& \\
\hline
\multicolumn{2}{l}{\emph{Planetary parameters}} \\
\hline
$M_{\rm p,b} ~[\rm M_\oplus]$  & $5.28\pm 0.77 $ \\
$R_{\rm p,b} ~[\rm R_\oplus]$  & $1.77\pm 0.07$ \\
$\rho_{\rm p,b}$ [$\rm g\;cm^{-3}$] & $5.3\pm 1.0 $ \\
$a_b$ [AU] & $0.0564\pm0.0012 $ \\
$M_{\rm p,c}\sin i ~[\rm M_\oplus]$  & $11.25\pm 1.1 $ \\
$a_c$ [AU] & $0.1155\pm0.0025 $ \\
\hline       
\hline
\vspace{-0.3cm}
\end{tabular}
\end{table}

\begin{figure}
    \includegraphics[width=\textwidth]{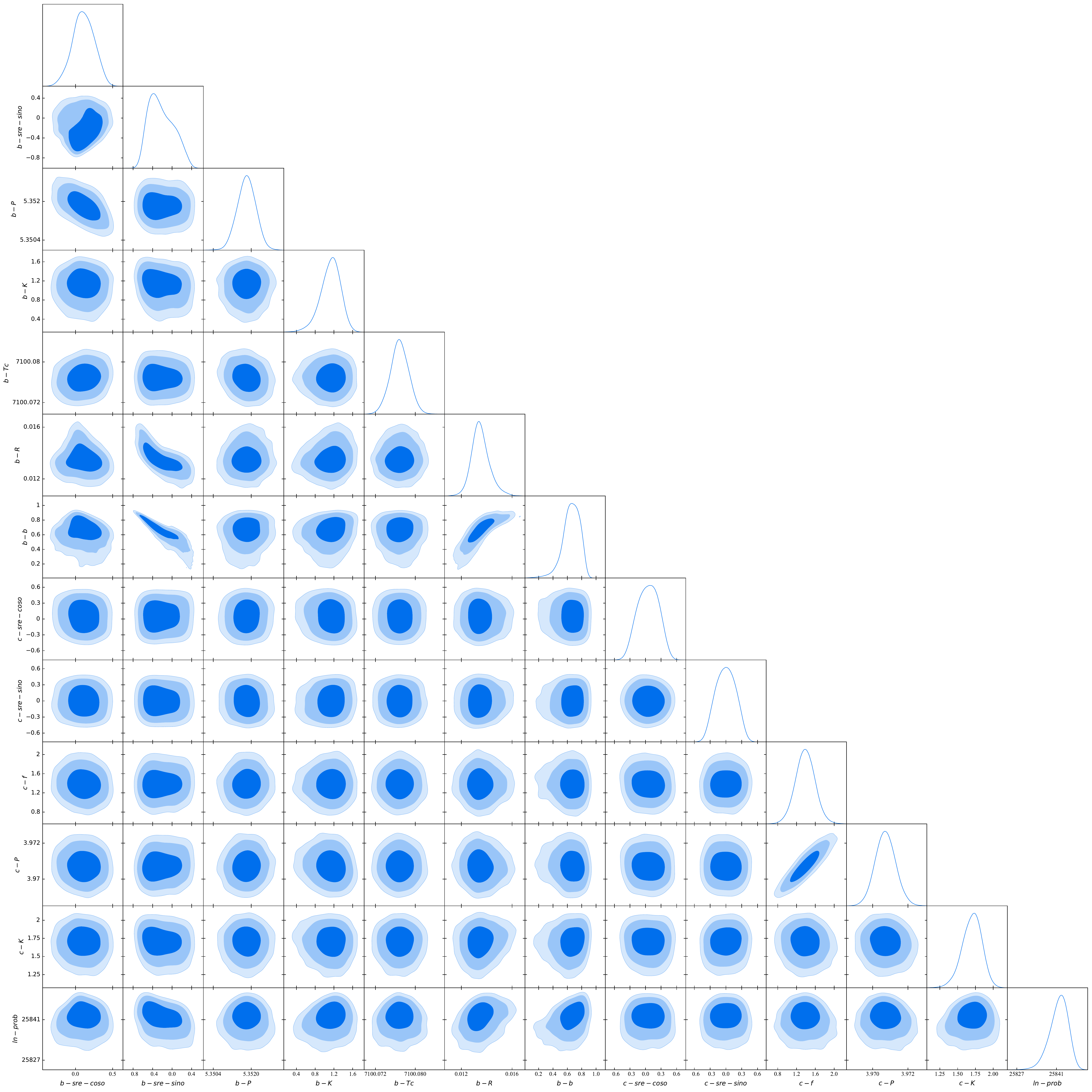}
    \caption{Corner plot of the planetary fitting parameters of the adopted two Keplerian model.}
    \label{fig:corner}
\end{figure}

\begin{figure}
    \includegraphics[width=\textwidth]{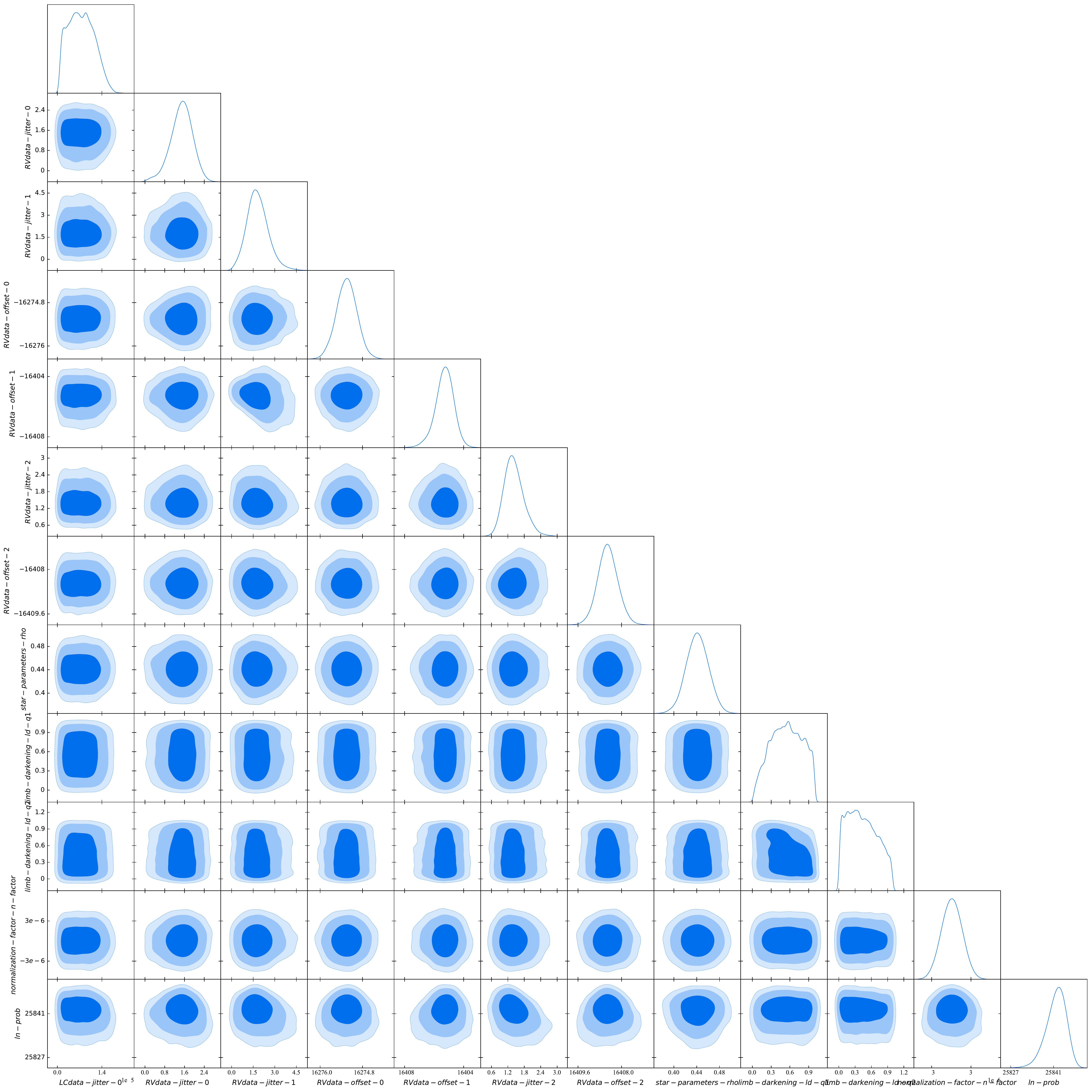}
    \caption{Corner plot of the stellar and instrumental fitting parameters of the adopted two Keplerian model.}
    \label{fig:corner2}
\end{figure}

\begin{figure}
    \includegraphics[width=\linewidth]{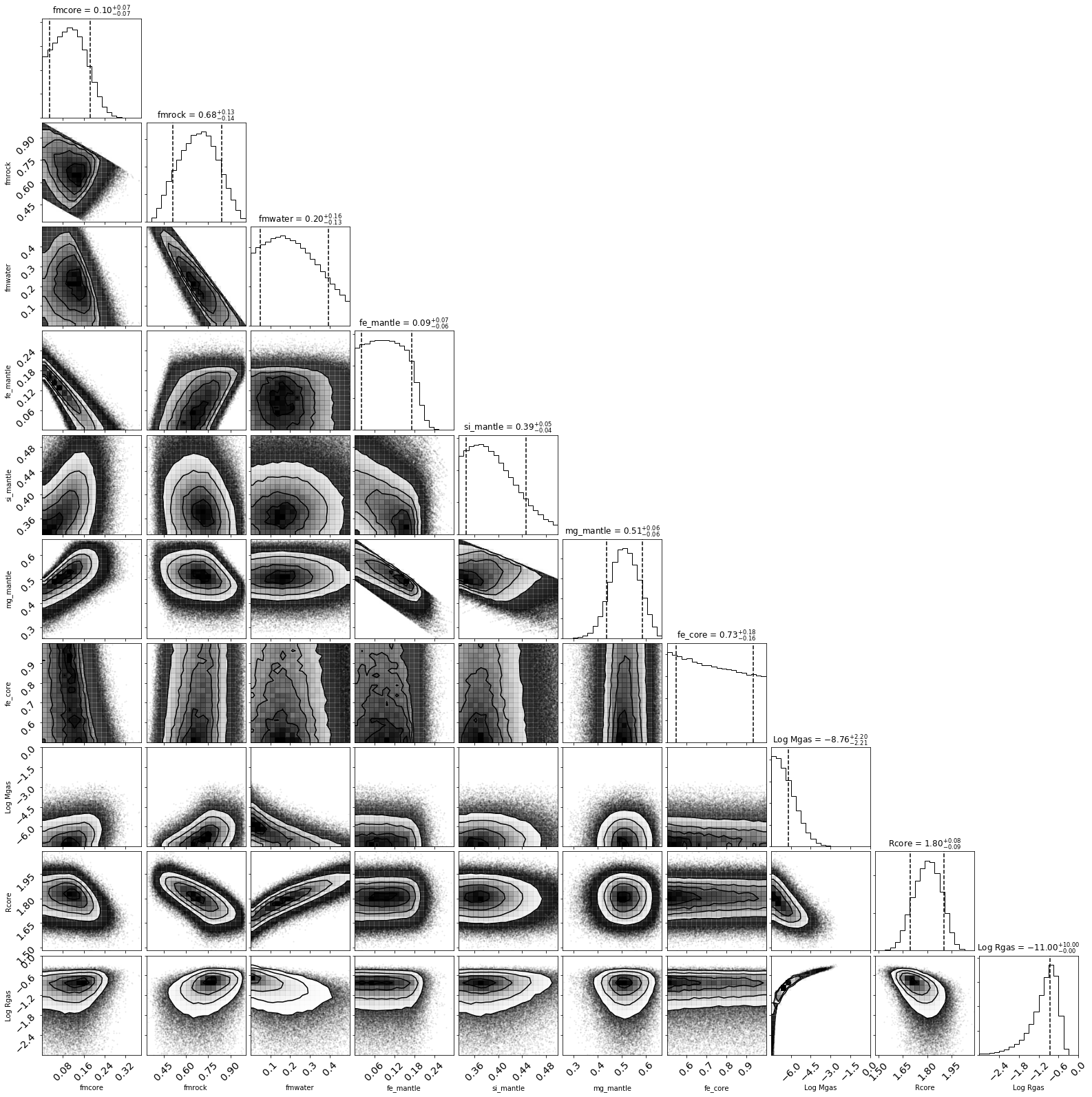}
    \caption{Corner plot of the interior structure parameters for the model including a gas envelope and a water layer. The mean and 11\% and 89\% quantiles are indicated on top of each histogram on the diagonal. fmcore is the mass fraction of the iron core, fmrock the fraction of silicate mantle, fmwater the fraction of water (all relative to the solid part of the planet). fe\_mantle, si\_mantle and mg\_mantle are the mole ratio of Fe, Si and Mg in the silicate mantle, fe\_core is the mole ratio of iron in the iron core (which contains iron and sulfur), Mgas and Rgas are the mass and thickness of the gas layer (in Earth units), and Rcore is the radius of the solid (iron, rock and water) core.}
    \label{fig:interior}
\end{figure}

\begin{figure}
    \includegraphics[width=\linewidth]{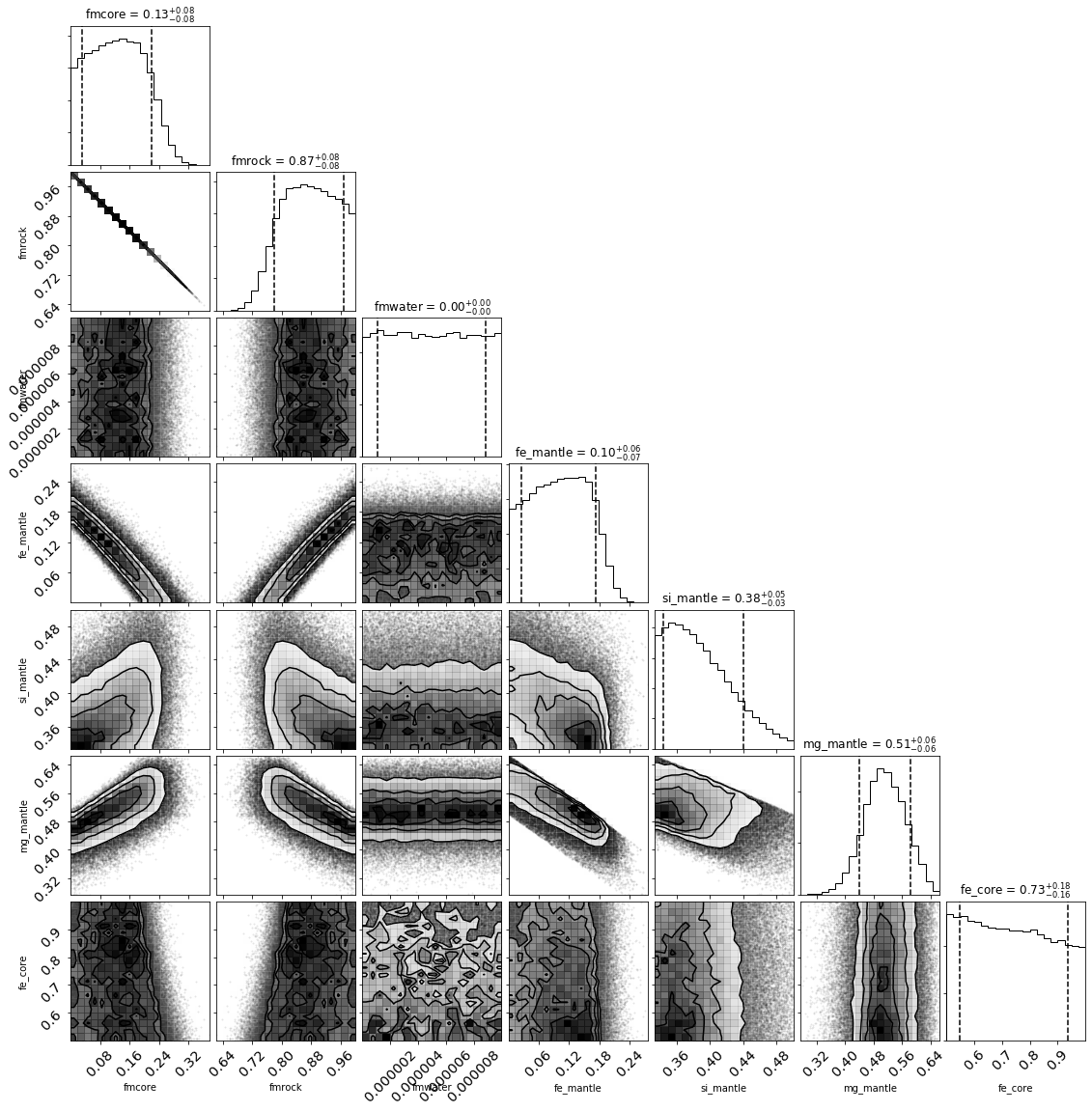}
    \caption{Corner plot of the interior structure parameters for the model with only an iron core and a silicate mantle. The mean and 11\% and 89\% quantiles are indicated on top of each histogram on the diagonal. Quantities are the same as in Figure \ref{fig:interior}.}
    \label{fig:interior2}
\end{figure}

\begin{landscape}
\begin{longtable}{llllllllllllll}
\caption{Radial velocities and activity indicators for \thisstar. The full table can also be found online.}            
\\\hline\hline
Source & Time & RV & $\sigma_{\text{RV}}$ & FWHM & $\sigma_{\text{FWHM}}$ & H$\alpha$ & $\sigma_{\text{H}\alpha}$ & NaID & $\sigma_{\text{NaID}}$ & S$_{\text{MW}}$ & $\sigma_{\text{S}}$ & $\log{R'_{HK}}$ & $\sigma_{\text{RHK}}$ \\
 & {[}BJD-2450000] & [km\,s$^{-1}$] & [km\,s$^{-1}$] & [km\,s$^{-1}$] & [km\,s$^{-1}$] & \\
\hline \endfirsthead
\caption[]{Cont. }
\\\hline\hline
Source & Time & RV & $\sigma_{\text{RV}}$ & FWHM & $\sigma_{\text{FWHM}}$ & H$\alpha$ & $\sigma_{\text{H}\alpha}$ & NaID & $\sigma_{\text{NaID}}$ & S$_{\text{MW}}$ & $\sigma_{\text{S}}$ & $\log{R'_{HK}}$ & $\sigma_{\text{RHK}}$ \\
 & {[}BJD-2450000] & [km\,s$^{-1}$] & [km\,s$^{-1}$] & [km\,s$^{-1}$] & [km\,s$^{-1}$] & \\
\hline \endhead
\hline\hline \endfoot
HARPS-N & $7321.7177$ & $-16.2835$ & $0.0025$ & $6.6726$ & $0.0050$ & $0.3177$ & $0.0004$ & $0.5475$ & $0.0003$ & $0.1637$ & $0.0005$ & $-4.9771$ & $0.0032$\\
HARPS-N & $7323.5753$ & $-16.2785$ & $0.0024$ & $6.6673$ & $0.0049$ & $0.3220$ & $0.0005$ & $0.5553$ & $0.0003$ & $0.1732$ & $0.0005$ & $-4.9246$ & $0.0028$\\
HARPS-N & $7324.6572$ & $-16.2826$ & $0.0027$ & $6.6753$ & $0.0054$ & $0.3150$ & $0.0005$ & $0.5604$ & $0.0003$ & $0.1768$ & $0.0006$ & $-4.9064$ & $0.0030$\\
HARPS-N & $7325.6206$ & $-16.2790$ & $0.0022$ & $6.6739$ & $0.0043$ & $0.3134$ & $0.0003$ & $0.5522$ & $0.0003$ & $0.1646$ & $0.0005$ & $-4.9715$ & $0.0027$\\
HARPS-N & $7330.5737$ & $-16.2723$ & $0.0021$ & $6.6701$ & $0.0042$ & $0.3172$ & $0.0004$ & $0.5562$ & $0.0003$ & $0.1710$ & $0.0005$ & $-4.9361$ & $0.0024$\\
HARPS-N & $7331.6683$ & $-16.2741$ & $0.0021$ & $6.6674$ & $0.0042$ & $0.3180$ & $0.0003$ & $0.5553$ & $0.0003$ & $0.1716$ & $0.0005$ & $-4.9330$ & $0.0025$\\
HARPS-N & $7332.6400$ & $-16.2708$ & $0.0027$ & $6.6726$ & $0.0055$ & $0.3166$ & $0.0005$ & $0.5644$ & $0.0004$ & $0.1715$ & $0.0006$ & $-4.9336$ & $0.0033$\\
HARPS-N & $7333.6249$ & $-16.2733$ & $0.0022$ & $6.6648$ & $0.0043$ & $0.3189$ & $0.0005$ & $0.5739$ & $0.0003$ & $0.1702$ & $0.0005$ & $-4.9403$ & $0.0026$\\
HARPS-N & $7334.5165$ & $-16.2711$ & $0.0022$ & $6.6723$ & $0.0045$ & $0.3168$ & $0.0004$ & $0.5594$ & $0.0003$ & $0.1808$ & $0.0005$ & $-4.8868$ & $0.0025$\\
HARPS-N & $7336.4903$ & $-16.2691$ & $0.0071$ & $6.6762$ & $0.0141$ & $0.3203$ & $0.0013$ & $0.7879$ & $0.0011$ & $0.1709$ & $0.0018$ & $-4.9368$ & $0.0097$\\
HARPS-N & $7345.5655$ & $-16.2657$ & $0.0038$ & $6.6586$ & $0.0075$ & $0.3186$ & $0.0007$ & $0.5934$ & $0.0005$ & $0.1644$ & $0.0009$ & $-4.9729$ & $0.0052$\\
HARPS-N & $7345.5917$ & $-16.2661$ & $0.0043$ & $6.6695$ & $0.0086$ & $0.3192$ & $0.0009$ & $0.6916$ & $0.0007$ & $0.1829$ & $0.0011$ & $-4.8766$ & $0.0050$\\
HARPS-N & $7345.6099$ & $-16.2694$ & $0.0045$ & $6.6711$ & $0.0091$ & $0.3317$ & $0.0009$ & $0.7293$ & $0.0007$ & $0.1566$ & $0.0011$ & $-5.0211$ & $0.0069$\\
HARPS-N & $7346.5838$ & $-16.2680$ & $0.0087$ & $6.6400$ & $0.0174$ & $0.3311$ & $0.0015$ & $0.6924$ & $0.0013$ & $0.1441$ & $0.0021$ & $-5.1106$ & $0.0171$\\
HARPS-N & $7370.5407$ & $-16.2753$ & $0.0025$ & $6.6697$ & $0.0050$ & $0.3265$ & $0.0005$ & $0.5569$ & $0.0004$ & $0.1729$ & $0.0006$ & $-4.9260$ & $0.0031$\\
HARPS-N & $7370.5616$ & $-16.2766$ & $0.0026$ & $6.6784$ & $0.0052$ & $0.3222$ & $0.0005$ & $0.5533$ & $0.0004$ & $0.1682$ & $0.0006$ & $-4.9515$ & $0.0033$\\
HARPS-N & $7371.4575$ & $-16.2750$ & $0.0026$ & $6.6686$ & $0.0051$ & $0.3239$ & $0.0005$ & $0.5679$ & $0.0003$ & $0.1644$ & $0.0006$ & $-4.9729$ & $0.0034$\\
HARPS-N & $7371.4788$ & $-16.2787$ & $0.0020$ & $6.6700$ & $0.0040$ & $0.3228$ & $0.0004$ & $0.5551$ & $0.0003$ & $0.1646$ & $0.0004$ & $-4.9717$ & $0.0025$\\
HARPS-N & $7374.5607$ & $-16.2784$ & $0.0025$ & $6.6673$ & $0.0050$ & $0.3135$ & $0.0004$ & $0.5475$ & $0.0003$ & $0.1695$ & $0.0006$ & $-4.9443$ & $0.0031$\\
HARPS-N & $7379.5495$ & $-16.2700$ & $0.0268$ & $6.7111$ & $0.0536$ & $0.3301$ & $0.0042$ & $0.5540$ & $0.0032$ & $0.1961$ & $0.0072$ & $-4.8194$ & $0.0292$\\
HARPS-N & $7380.5195$ & $-16.2803$ & $0.0036$ & $6.6629$ & $0.0073$ & $0.3132$ & $0.0006$ & $0.5502$ & $0.0005$ & $0.1774$ & $0.0009$ & $-4.9032$ & $0.0044$\\
HARPS-N & $7381.5605$ & $-16.2737$ & $0.0025$ & $6.6798$ & $0.0050$ & $0.3250$ & $0.0005$ & $0.5544$ & $0.0003$ & $0.1586$ & $0.0005$ & $-5.0080$ & $0.0035$\\
HARPS-N & $7384.5048$ & $-16.2798$ & $0.0027$ & $6.6729$ & $0.0054$ & $0.3229$ & $0.0005$ & $0.5500$ & $0.0004$ & $0.1724$ & $0.0006$ & $-4.9285$ & $0.0033$\\
HARPS-N & $7399.3231$ & $-16.2740$ & $0.0054$ & $6.6728$ & $0.0108$ & $0.3148$ & $0.0013$ & $0.5706$ & $0.0009$ & $0.1529$ & $0.0013$ & $-5.0458$ & $0.0089$\\
HARPS-N & $7407.3349$ & $-16.2739$ & $0.0028$ & $6.7004$ & $0.0056$ & $0.3219$ & $0.0005$ & $0.5496$ & $0.0004$ & $0.1717$ & $0.0006$ & $-4.9323$ & $0.0033$\\
HARPS-N & $7407.3569$ & $-16.2780$ & $0.0024$ & $6.6773$ & $0.0047$ & $0.3193$ & $0.0005$ & $0.5563$ & $0.0003$ & $0.1694$ & $0.0005$ & $-4.9448$ & $0.0029$\\
HARPS-N & $7408.3926$ & $-16.2688$ & $0.0031$ & $6.6700$ & $0.0063$ & $0.3214$ & $0.0006$ & $0.5537$ & $0.0004$ & $0.1716$ & $0.0007$ & $-4.9328$ & $0.0039$\\
HARPS-N & $7408.4146$ & $-16.2674$ & $0.0037$ & $6.6780$ & $0.0074$ & $0.3222$ & $0.0007$ & $0.5471$ & $0.0005$ & $0.1733$ & $0.0009$ & $-4.9242$ & $0.0046$\\
HARPS-N & $7409.3495$ & $-16.2740$ & $0.0026$ & $6.6777$ & $0.0051$ & $0.3241$ & $0.0005$ & $0.5520$ & $0.0003$ & $0.1628$ & $0.0006$ & $-4.9825$ & $0.0034$\\
HARPS-N & $7409.3696$ & $-16.2758$ & $0.0042$ & $6.6561$ & $0.0084$ & $0.3285$ & $0.0008$ & $0.5577$ & $0.0006$ & $0.1827$ & $0.0010$ & $-4.8778$ & $0.0045$\\
HARPS-N & $7411.3837$ & $-16.2711$ & $0.0046$ & $6.6566$ & $0.0092$ & $0.3019$ & $0.0005$ & $0.5262$ & $0.0005$ & $0.1961$ & $0.0011$ & $-4.8191$ & $0.0044$\\
HARPS-N & $7411.4046$ & $-16.2690$ & $0.0053$ & $6.6748$ & $0.0107$ & $0.3022$ & $0.0005$ & $0.5115$ & $0.0006$ & $0.1767$ & $0.0012$ & $-4.9068$ & $0.0058$\\
HARPS-N & $7412.4078$ & $-16.2747$ & $0.0026$ & $6.6842$ & $0.0053$ & $0.3167$ & $0.0004$ & $0.5431$ & $0.0003$ & $0.1630$ & $0.0006$ & $-4.9813$ & $0.0035$\\
HARPS-N & $7412.4296$ & $-16.2743$ & $0.0024$ & $6.6691$ & $0.0047$ & $0.3151$ & $0.0004$ & $0.5469$ & $0.0003$ & $0.1768$ & $0.0005$ & $-4.9059$ & $0.0026$\\
HARPS-N & $7413.3898$ & $-16.2749$ & $0.0028$ & $6.6751$ & $0.0057$ & $0.3158$ & $0.0004$ & $0.5461$ & $0.0004$ & $0.1617$ & $0.0006$ & $-4.9889$ & $0.0039$\\
HARPS-N & $7413.4108$ & $-16.2690$ & $0.0027$ & $6.6818$ & $0.0055$ & $0.3154$ & $0.0004$ & $0.5470$ & $0.0003$ & $0.1674$ & $0.0006$ & $-4.9560$ & $0.0034$\\
HARPS-N & $7414.3786$ & $-16.2722$ & $0.0022$ & $6.6798$ & $0.0044$ & $0.3220$ & $0.0004$ & $0.5548$ & $0.0003$ & $0.1647$ & $0.0005$ & $-4.9712$ & $0.0027$\\
HARPS-N & $7414.3990$ & $-16.2768$ & $0.0025$ & $6.6695$ & $0.0049$ & $0.3216$ & $0.0004$ & $0.5500$ & $0.0003$ & $0.1710$ & $0.0006$ & $-4.9360$ & $0.0030$\\
HARPS-N & $7415.4532$ & $-16.2789$ & $0.0024$ & $6.6622$ & $0.0048$ & $0.3172$ & $0.0004$ & $0.5438$ & $0.0003$ & $0.1724$ & $0.0005$ & $-4.9288$ & $0.0028$\\
HARPS-N & $7415.4717$ & $-16.2754$ & $0.0034$ & $6.6655$ & $0.0068$ & $0.3147$ & $0.0006$ & $0.5481$ & $0.0004$ & $0.1746$ & $0.0008$ & $-4.9171$ & $0.0041$\\
HARPS-N & $7432.3624$ & $-16.2816$ & $0.0027$ & $6.6723$ & $0.0054$ & $0.3162$ & $0.0004$ & $0.5371$ & $0.0003$ & $0.1692$ & $0.0006$ & $-4.9457$ & $0.0032$\\
HARPS-N & $7432.3839$ & $-16.2814$ & $0.0025$ & $6.6668$ & $0.0050$ & $0.3081$ & $0.0003$ & $0.5406$ & $0.0003$ & $0.1632$ & $0.0005$ & $-4.9802$ & $0.0032$\\
HARPS-N & $7433.3715$ & $-16.2747$ & $0.0030$ & $6.6679$ & $0.0060$ & $0.3237$ & $0.0005$ & $0.5456$ & $0.0004$ & $0.1812$ & $0.0007$ & $-4.8846$ & $0.0033$\\
HARPS-N & $7433.3943$ & $-16.2760$ & $0.0034$ & $6.6886$ & $0.0068$ & $0.3241$ & $0.0006$ & $0.5510$ & $0.0004$ & $0.1741$ & $0.0008$ & $-4.9197$ & $0.0040$\\
HARPS-N & $7435.4321$ & $-16.2782$ & $0.0038$ & $6.6720$ & $0.0076$ & $0.3111$ & $0.0005$ & $0.5382$ & $0.0004$ & $0.1923$ & $0.0009$ & $-4.8349$ & $0.0039$\\
HARPS-N & $7436.3951$ & $-16.2774$ & $0.0026$ & $6.6640$ & $0.0052$ & $0.3213$ & $0.0006$ & $0.5495$ & $0.0004$ & $0.1714$ & $0.0006$ & $-4.9339$ & $0.0031$\\
HARPS-N & $7436.4171$ & $-16.2794$ & $0.0023$ & $6.6637$ & $0.0047$ & $0.3203$ & $0.0006$ & $0.5532$ & $0.0003$ & $0.1643$ & $0.0005$ & $-4.9732$ & $0.0030$\\
HARPS-N & $7458.4045$ & $-16.2687$ & $0.0029$ & $6.6805$ & $0.0057$ & $0.3208$ & $0.0005$ & $0.5457$ & $0.0004$ & $0.1756$ & $0.0007$ & $-4.9120$ & $0.0035$\\
HARPS-N & $7458.4254$ & $-16.2703$ & $0.0030$ & $6.6812$ & $0.0060$ & $0.3160$ & $0.0005$ & $0.5453$ & $0.0004$ & $0.1765$ & $0.0007$ & $-4.9076$ & $0.0037$\\
HARPS-N & $7461.3722$ & $-16.2761$ & $0.0020$ & $6.6722$ & $0.0041$ & $0.3143$ & $0.0003$ & $0.5524$ & $0.0002$ & $0.1568$ & $0.0004$ & $-5.0197$ & $0.0027$\\
HARPS-N & $7461.3929$ & $-16.2776$ & $0.0022$ & $6.6775$ & $0.0045$ & $0.3125$ & $0.0003$ & $0.5454$ & $0.0003$ & $0.1703$ & $0.0005$ & $-4.9399$ & $0.0026$\\
HARPS-N & $7462.3885$ & $-16.2773$ & $0.0023$ & $6.6690$ & $0.0046$ & $0.3204$ & $0.0004$ & $0.5540$ & $0.0003$ & $0.1683$ & $0.0005$ & $-4.9510$ & $0.0028$\\
HARPS-N & $7462.4088$ & $-16.2751$ & $0.0022$ & $6.6682$ & $0.0044$ & $0.3169$ & $0.0004$ & $0.5503$ & $0.0003$ & $0.1630$ & $0.0005$ & $-4.9813$ & $0.0030$\\
HARPS-N & $7463.4056$ & $-16.2788$ & $0.0023$ & $6.6701$ & $0.0046$ & $0.3138$ & $0.0004$ & $0.5468$ & $0.0003$ & $0.1647$ & $0.0005$ & $-4.9713$ & $0.0030$\\
HARPS-N & $7463.4261$ & $-16.2783$ & $0.0027$ & $6.6785$ & $0.0054$ & $0.3147$ & $0.0004$ & $0.5443$ & $0.0003$ & $0.1767$ & $0.0007$ & $-4.9068$ & $0.0033$\\
HARPS-N & $8142.4204$ & $-16.2703$ & $0.0023$ & $6.6844$ & $0.0045$ & $0.3196$ & $0.0004$ & $0.5498$ & $0.0003$ & $0.1747$ & $0.0005$ & $-4.9169$ & $0.0026$\\
HARPS-N & $8144.4449$ & $-16.2795$ & $0.0023$ & $6.6778$ & $0.0046$ & $0.3271$ & $0.0005$ & $0.5544$ & $0.0003$ & $0.1660$ & $0.0007$ & $-4.9637$ & $0.0040$\\
HARPS-N & $8145.3940$ & $-16.2749$ & $0.0028$ & $6.6841$ & $0.0057$ & $0.3265$ & $0.0006$ & $0.5514$ & $0.0004$ & $0.1667$ & $0.0006$ & $-4.9597$ & $0.0035$\\
HARPS-N & $8145.5051$ & $-16.2759$ & $0.0043$ & $6.6898$ & $0.0085$ & $0.3295$ & $0.0008$ & $0.5628$ & $0.0005$ & $0.2098$ & $0.0034$ & $-4.7664$ & $0.0123$\\
HARPS-N & $8172.4232$ & $-16.2824$ & $0.0036$ & $6.6837$ & $0.0072$ & $0.3112$ & $0.0005$ & $0.5434$ & $0.0004$ & $0.1629$ & $0.0008$ & $-4.9816$ & $0.0047$\\
HARPS-N & $8184.3940$ & $-16.2734$ & $0.0025$ & $6.6781$ & $0.0050$ & $0.3124$ & $0.0003$ & $0.5321$ & $0.0003$ & $0.1646$ & $0.0005$ & $-4.9718$ & $0.0030$\\
HARPS-N & $8188.4256$ & $-16.2799$ & $0.0057$ & $6.6668$ & $0.0115$ & $0.3192$ & $0.0009$ & $0.5473$ & $0.0007$ & $0.1601$ & $0.0013$ & $-4.9984$ & $0.0079$\\
HARPS-N & $8189.3954$ & $-16.2709$ & $0.0035$ & $6.6675$ & $0.0069$ & $0.3094$ & $0.0004$ & $0.5238$ & $0.0004$ & $0.1545$ & $0.0008$ & $-5.0346$ & $0.0052$\\
HARPS-N & $8192.4030$ & $-16.2708$ & $0.0042$ & $6.6904$ & $0.0084$ & $0.3121$ & $0.0006$ & $0.5340$ & $0.0005$ & $0.1879$ & $0.0009$ & $-4.8541$ & $0.0042$\\
HARPS-N & $8360.6795$ & $-16.2799$ & $0.0035$ & $6.6830$ & $0.0069$ & $0.3177$ & $0.0005$ & $0.5444$ & $0.0004$ & $0.1687$ & $0.0007$ & $-4.9488$ & $0.0041$\\
HARPS-N & $8361.6623$ & $-16.2771$ & $0.0070$ & $6.6918$ & $0.0141$ & $0.3178$ & $0.0010$ & $0.5484$ & $0.0009$ & $0.2122$ & $0.0017$ & $-4.7578$ & $0.0061$\\
HARPS-N & $8364.6554$ & $-16.2773$ & $0.0019$ & $6.6759$ & $0.0038$ & $0.3199$ & $0.0004$ & $0.5510$ & $0.0002$ & $0.1664$ & $0.0004$ & $-4.9613$ & $0.0021$\\
HARPS-N & $8365.6759$ & $-16.2728$ & $0.0022$ & $6.6839$ & $0.0044$ & $0.3216$ & $0.0004$ & $0.5515$ & $0.0003$ & $0.1711$ & $0.0004$ & $-4.9354$ & $0.0023$\\
HARPS-N & $8366.7071$ & $-16.2733$ & $0.0022$ & $6.6860$ & $0.0044$ & $0.3287$ & $0.0005$ & $0.5520$ & $0.0003$ & $0.1604$ & $0.0004$ & $-4.9965$ & $0.0027$\\
HARPS-N & $8378.6900$ & $-16.2734$ & $0.0023$ & $6.6736$ & $0.0046$ & $0.3165$ & $0.0004$ & $0.5468$ & $0.0003$ & $0.1689$ & $0.0005$ & $-4.9475$ & $0.0026$\\
HARPS-N & $8379.6411$ & $-16.2749$ & $0.0024$ & $6.6753$ & $0.0048$ & $0.3164$ & $0.0004$ & $0.5461$ & $0.0003$ & $0.1771$ & $0.0005$ & $-4.9046$ & $0.0025$\\
HARPS-N & $8380.6361$ & $-16.2732$ & $0.0017$ & $6.6739$ & $0.0034$ & $0.3203$ & $0.0003$ & $0.5480$ & $0.0002$ & $0.1678$ & $0.0003$ & $-4.9536$ & $0.0017$\\
HARPS-N & $8381.6411$ & $-16.2748$ & $0.0017$ & $6.6715$ & $0.0034$ & $0.3219$ & $0.0003$ & $0.5535$ & $0.0002$ & $0.1653$ & $0.0003$ & $-4.9679$ & $0.0018$\\
HARPS-N & $8382.6416$ & $-16.2682$ & $0.0026$ & $6.6793$ & $0.0052$ & $0.3171$ & $0.0005$ & $0.5409$ & $0.0003$ & $0.1670$ & $0.0005$ & $-4.9578$ & $0.0031$\\
HARPS-N & $8383.6109$ & $-16.2819$ & $0.0049$ & $6.6616$ & $0.0097$ & $0.3164$ & $0.0007$ & $0.5374$ & $0.0006$ & $0.1756$ & $0.0011$ & $-4.9122$ & $0.0058$\\
HARPS-N & $8384.6447$ & $-16.2740$ & $0.0039$ & $6.6892$ & $0.0079$ & $0.3157$ & $0.0006$ & $0.5375$ & $0.0005$ & $0.1622$ & $0.0009$ & $-4.9860$ & $0.0052$\\
HARPS-N & $8385.6476$ & $-16.2744$ & $0.0021$ & $6.6772$ & $0.0042$ & $0.3221$ & $0.0004$ & $0.5510$ & $0.0003$ & $0.1694$ & $0.0004$ & $-4.9448$ & $0.0023$\\
HARPS-N & $8386.6543$ & $-16.2750$ & $0.0019$ & $6.6733$ & $0.0038$ & $0.3179$ & $0.0003$ & $0.5433$ & $0.0002$ & $0.1668$ & $0.0004$ & $-4.9593$ & $0.0020$\\
HARPS-N & $8388.6768$ & $-16.2763$ & $0.0023$ & $6.6715$ & $0.0046$ & $0.3160$ & $0.0004$ & $0.5477$ & $0.0003$ & $0.1558$ & $0.0004$ & $-5.0262$ & $0.0029$\\
HARPS-N & $8390.7365$ & $-16.2811$ & $0.0020$ & $6.6793$ & $0.0040$ & $0.3186$ & $0.0003$ & $0.5486$ & $0.0002$ & $0.1711$ & $0.0004$ & $-4.9355$ & $0.0021$\\
HARPS-N & $8391.7053$ & $-16.2762$ & $0.0021$ & $6.6858$ & $0.0043$ & $0.3197$ & $0.0003$ & $0.5484$ & $0.0003$ & $0.1725$ & $0.0004$ & $-4.9279$ & $0.0022$\\
HARPS-N & $8415.6638$ & $-16.2751$ & $0.0026$ & $6.6758$ & $0.0052$ & $0.3147$ & $0.0004$ & $0.5407$ & $0.0003$ & $0.1674$ & $0.0005$ & $-4.9556$ & $0.0030$\\
HARPS-N & $8448.6711$ & $-16.2772$ & $0.0022$ & $6.6789$ & $0.0044$ & $0.3148$ & $0.0003$ & $0.5525$ & $0.0003$ & $0.1632$ & $0.0004$ & $-4.9801$ & $0.0026$\\
HARPS-N & $8451.4508$ & $-16.2756$ & $0.0024$ & $6.6738$ & $0.0048$ & $0.3220$ & $0.0005$ & $0.5667$ & $0.0003$ & $0.1722$ & $0.0005$ & $-4.9295$ & $0.0026$\\
HARPS-N & $8453.6622$ & $-16.2724$ & $0.0022$ & $6.6857$ & $0.0045$ & $0.3154$ & $0.0003$ & $0.5461$ & $0.0003$ & $0.1681$ & $0.0005$ & $-4.9520$ & $0.0025$\\
HARPS-N & $8454.5250$ & $-16.2774$ & $0.0026$ & $6.6637$ & $0.0052$ & $0.3220$ & $0.0005$ & $0.5555$ & $0.0004$ & $0.1699$ & $0.0006$ & $-4.9419$ & $0.0030$\\
HARPS-N & $8456.4260$ & $-16.2781$ & $0.0030$ & $6.6736$ & $0.0060$ & $0.3173$ & $0.0005$ & $0.5572$ & $0.0004$ & $0.1776$ & $0.0007$ & $-4.9021$ & $0.0033$\\
HARPS-N & $8477.5414$ & $-16.2758$ & $0.0049$ & $6.6786$ & $0.0098$ & $0.3104$ & $0.0007$ & $0.5405$ & $0.0006$ & $0.1848$ & $0.0011$ & $-4.8680$ & $0.0052$\\
HARPS-N & $8478.6114$ & $-16.2772$ & $0.0025$ & $6.6734$ & $0.0050$ & $0.3192$ & $0.0005$ & $0.5451$ & $0.0003$ & $0.1677$ & $0.0005$ & $-4.9541$ & $0.0030$\\
HARPS-N & $8480.5598$ & $-16.2743$ & $0.0021$ & $6.6754$ & $0.0041$ & $0.3269$ & $0.0005$ & $0.5550$ & $0.0003$ & $0.1622$ & $0.0004$ & $-4.9860$ & $0.0024$\\
HARPS-N & $8481.5445$ & $-16.2798$ & $0.0026$ & $6.6776$ & $0.0052$ & $0.3171$ & $0.0005$ & $0.5463$ & $0.0003$ & $0.1729$ & $0.0006$ & $-4.9263$ & $0.0029$\\
HARPS-N & $8482.5003$ & $-16.2783$ & $0.0020$ & $6.6848$ & $0.0041$ & $0.3299$ & $0.0006$ & $0.5584$ & $0.0003$ & $0.1732$ & $0.0004$ & $-4.9245$ & $0.0021$\\
HARPS-N & $8483.5377$ & $-16.2828$ & $0.0022$ & $6.6694$ & $0.0043$ & $0.3252$ & $0.0005$ & $0.5560$ & $0.0003$ & $0.1760$ & $0.0004$ & $-4.9100$ & $0.0022$\\
HARPS-N & $8484.5243$ & $-16.2764$ & $0.0022$ & $6.6889$ & $0.0044$ & $0.3244$ & $0.0004$ & $0.5546$ & $0.0003$ & $0.1687$ & $0.0004$ & $-4.9488$ & $0.0025$\\
HARPS-N & $8487.4733$ & $-16.2820$ & $0.0023$ & $6.6719$ & $0.0046$ & $0.3168$ & $0.0004$ & $0.5456$ & $0.0003$ & $0.1654$ & $0.0005$ & $-4.9671$ & $0.0027$\\
HARPS-N & $8488.4399$ & $-16.2801$ & $0.0020$ & $6.6827$ & $0.0039$ & $0.3270$ & $0.0003$ & $0.5330$ & $0.0002$ & $0.1722$ & $0.0004$ & $-4.9298$ & $0.0020$\\
HARPS-N & $8489.4970$ & $-16.2742$ & $0.0020$ & $6.6773$ & $0.0039$ & $0.3117$ & $0.0003$ & $0.5368$ & $0.0002$ & $0.1699$ & $0.0004$ & $-4.9418$ & $0.0020$\\
HARPS-N & $8502.4118$ & $-16.2733$ & $0.0028$ & $6.6761$ & $0.0055$ & $0.3160$ & $0.0004$ & $0.5409$ & $0.0003$ & $0.1776$ & $0.0006$ & $-4.9024$ & $0.0029$\\
HARPS-N & $8504.4724$ & $-16.2728$ & $0.0026$ & $6.6749$ & $0.0052$ & $0.3106$ & $0.0003$ & $0.5271$ & $0.0003$ & $0.1640$ & $0.0005$ & $-4.9750$ & $0.0030$\\
HARPS-N & $8505.4046$ & $-16.2646$ & $0.0026$ & $6.6923$ & $0.0052$ & $0.3132$ & $0.0003$ & $0.5389$ & $0.0003$ & $0.1725$ & $0.0005$ & $-4.9282$ & $0.0028$\\
HARPS-N & $8518.3509$ & $-16.2691$ & $0.0028$ & $6.6834$ & $0.0057$ & $0.3149$ & $0.0005$ & $0.5385$ & $0.0003$ & $0.1790$ & $0.0006$ & $-4.8955$ & $0.0030$\\
HARPS-N & $8520.3998$ & $-16.2734$ & $0.0023$ & $6.6768$ & $0.0047$ & $0.3170$ & $0.0004$ & $0.5445$ & $0.0003$ & $0.1752$ & $0.0005$ & $-4.9140$ & $0.0025$\\
HARPS-N & $8521.3533$ & $-16.2663$ & $0.0028$ & $6.6787$ & $0.0055$ & $0.3232$ & $0.0005$ & $0.5404$ & $0.0004$ & $0.1676$ & $0.0006$ & $-4.9544$ & $0.0032$\\
HARPS-N & $8522.3907$ & $-16.2708$ & $0.0022$ & $6.6817$ & $0.0045$ & $0.3212$ & $0.0004$ & $0.5521$ & $0.0003$ & $0.1652$ & $0.0004$ & $-4.9681$ & $0.0025$\\
HARPS-N & $8523.4821$ & $-16.2697$ & $0.0021$ & $6.6792$ & $0.0042$ & $0.3161$ & $0.0003$ & $0.5476$ & $0.0002$ & $0.1687$ & $0.0004$ & $-4.9485$ & $0.0024$\\
HARPS-N & $8524.3584$ & $-16.2717$ & $0.0021$ & $6.6853$ & $0.0043$ & $0.3190$ & $0.0004$ & $0.5498$ & $0.0003$ & $0.1657$ & $0.0004$ & $-4.9653$ & $0.0024$\\
HARPS-N & $8525.3566$ & $-16.2785$ & $0.0031$ & $6.6795$ & $0.0062$ & $0.3169$ & $0.0006$ & $0.5499$ & $0.0004$ & $0.1725$ & $0.0007$ & $-4.9283$ & $0.0035$\\
HARPS-N & $8526.3852$ & $-16.2712$ & $0.0025$ & $6.6690$ & $0.0051$ & $0.3190$ & $0.0005$ & $0.5470$ & $0.0003$ & $0.1724$ & $0.0005$ & $-4.9287$ & $0.0027$\\
HARPS-N & $8527.3507$ & $-16.2697$ & $0.0022$ & $6.6801$ & $0.0044$ & $0.3276$ & $0.0005$ & $0.5520$ & $0.0003$ & $0.1627$ & $0.0004$ & $-4.9830$ & $0.0026$\\
HARPS-N & $8528.3547$ & $-16.2736$ & $0.0029$ & $6.6830$ & $0.0057$ & $0.3209$ & $0.0006$ & $0.5502$ & $0.0004$ & $0.1684$ & $0.0006$ & $-4.9503$ & $0.0032$\\
HARPS-N & $8537.3930$ & $-16.2757$ & $0.0036$ & $6.6842$ & $0.0072$ & $0.3103$ & $0.0005$ & $0.5340$ & $0.0004$ & $0.1895$ & $0.0008$ & $-4.8468$ & $0.0036$\\
HARPS-N & $8538.4266$ & $-16.2647$ & $0.0032$ & $6.6730$ & $0.0064$ & $0.3112$ & $0.0005$ & $0.5364$ & $0.0004$ & $0.1632$ & $0.0007$ & $-4.9797$ & $0.0041$\\
HARPS-N & $8539.4355$ & $-16.2703$ & $0.0023$ & $6.6807$ & $0.0046$ & $0.3194$ & $0.0003$ & $0.5448$ & $0.0003$ & $0.1720$ & $0.0005$ & $-4.9306$ & $0.0025$\\
ESPRESSO & $8421.7397$ & $-16.4084$ & $0.0009$ & $6.8739$ & $0.0018$ & $0.3048$ & $0.0001$ & $0.5347$ & $0.0001$ & $0.1688$ & $0.0002$ & $-4.9478$ & $0.0011$\\
ESPRESSO & $8424.7793$ & $-16.4061$ & $0.0009$ & $6.8763$ & $0.0017$ & $0.3055$ & $0.0001$ & $0.5314$ & $0.0001$ & $0.1700$ & $0.0002$ & $-4.9416$ & $0.0010$\\
ESPRESSO & $8426.7576$ & $-16.4012$ & $0.0010$ & $6.8765$ & $0.0020$ & $0.3042$ & $0.0001$ & $0.5313$ & $0.0001$ & $0.1663$ & $0.0002$ & $-4.9618$ & $0.0014$\\
ESPRESSO & $8439.5931$ & $-16.4139$ & $0.0031$ & $6.8730$ & $0.0063$ & $0.3016$ & $0.0004$ & $0.5581$ & $0.0003$ & $0.1554$ & $0.0013$ & $-5.0285$ & $0.0087$\\
ESPRESSO & $8459.6339$ & $-16.4022$ & $0.0010$ & $6.8804$ & $0.0021$ & $0.3052$ & $0.0001$ & $0.5380$ & $0.0001$ & $0.1671$ & $0.0002$ & $-4.9575$ & $0.0014$\\
ESPRESSO & $8463.7307$ & $-16.4005$ & $0.0011$ & $6.8824$ & $0.0023$ & $0.3044$ & $0.0001$ & $0.5278$ & $0.0001$ & $0.1663$ & $0.0003$ & $-4.9621$ & $0.0017$\\
ESPRESSO & $8466.6280$ & $-16.4062$ & $0.0011$ & $6.8770$ & $0.0022$ & $0.3040$ & $0.0001$ & $0.5221$ & $0.0001$ & $0.1669$ & $0.0003$ & $-4.9586$ & $0.0014$\\
ESPRESSO & $8470.7001$ & $-16.4093$ & $0.0013$ & $6.8733$ & $0.0025$ & $0.3048$ & $0.0001$ & $0.5219$ & $0.0001$ & $0.1641$ & $0.0003$ & $-4.9747$ & $0.0020$\\
ESPRESSO & $8488.6354$ & $-16.4112$ & $0.0017$ & $6.8727$ & $0.0034$ & $0.2982$ & $0.0002$ & $0.5268$ & $0.0002$ & $0.1648$ & $0.0005$ & $-4.9708$ & $0.0029$\\
ESPRESSO & $8493.5928$ & $-16.4049$ & $0.0012$ & $6.8817$ & $0.0025$ & $0.3051$ & $0.0001$ & $0.5358$ & $0.0001$ & $0.1888$ & $0.0003$ & $-4.8500$ & $0.0014$\\
ESPRESSO & $8512.5379$ & $-16.4042$ & $0.0014$ & $6.8817$ & $0.0028$ & $0.2982$ & $0.0002$ & $0.5115$ & $0.0001$ & $0.1655$ & $0.0004$ & $-4.9667$ & $0.0023$\\
ESPRESSO & $8536.5397$ & $-16.4062$ & $0.0014$ & $6.8844$ & $0.0028$ & $0.3003$ & $0.0002$ & $0.5240$ & $0.0001$ & $0.1669$ & $0.0004$ & $-4.9584$ & $0.0024$\\
ESPRESSO & $8537.5597$ & $-16.4041$ & $0.0037$ & $6.8711$ & $0.0074$ & $0.2996$ & $0.0004$ & $0.5145$ & $0.0003$ & $0.1600$ & $0.0017$ & $-4.9991$ & $0.0104$\\
ESPRESSO & $8538.5237$ & $-16.3993$ & $0.0025$ & $6.8884$ & $0.0049$ & $0.3008$ & $0.0003$ & $0.5344$ & $0.0002$ & $0.1705$ & $0.0010$ & $-4.9387$ & $0.0051$\\
ESPRESSO & $8543.5334$ & $-16.4016$ & $0.0016$ & $6.8787$ & $0.0032$ & $0.2994$ & $0.0002$ & $0.5268$ & $0.0002$ & $0.1670$ & $0.0005$ & $-4.9581$ & $0.0030$\\
ESPRESSO & $8544.5259$ & $-16.4112$ & $0.0054$ & $6.8603$ & $0.0107$ & $0.2950$ & $0.0006$ & $0.5355$ & $0.0005$ & $0.1599$ & $0.0024$ & $-4.9998$ & $0.0149$\\
ESPRESSO & $8706.8772$ & $-16.4098$ & $0.0010$ & $6.8940$ & $0.0020$ & $0.3061$ & $0.0001$ & $0.5340$ & $0.0001$ & $0.1647$ & $0.0003$ & $-4.9709$ & $0.0015$\\
ESPRESSO & $8721.8890$ & $-16.4106$ & $0.0009$ & $6.8872$ & $0.0019$ & $0.3046$ & $0.0001$ & $0.5357$ & $0.0001$ & $0.1669$ & $0.0002$ & $-4.9586$ & $0.0014$\\
ESPRESSO & $8725.8671$ & $-16.4020$ & $0.0014$ & $6.8837$ & $0.0027$ & $0.3028$ & $0.0001$ & $0.5361$ & $0.0001$ & $0.1650$ & $0.0004$ & $-4.9691$ & $0.0025$\\
ESPRESSO & $8731.8553$ & $-16.4099$ & $0.0009$ & $6.8910$ & $0.0018$ & $0.3036$ & $0.0001$ & $0.5348$ & $0.0001$ & $0.1660$ & $0.0002$ & $-4.9635$ & $0.0013$\\
ESPRESSO & $8737.8864$ & $-16.4120$ & $0.0009$ & $6.8913$ & $0.0017$ & $0.3030$ & $0.0001$ & $0.5374$ & $0.0001$ & $0.1666$ & $0.0002$ & $-4.9605$ & $0.0011$\\
ESPRESSO & $8741.8028$ & $-16.4044$ & $0.0010$ & $6.8862$ & $0.0021$ & $0.3026$ & $0.0001$ & $0.5357$ & $0.0001$ & $0.1656$ & $0.0003$ & $-4.9660$ & $0.0016$\\
ESPRESSO & $8754.8140$ & $-16.4079$ & $0.0008$ & $6.8905$ & $0.0015$ & $0.3029$ & $0.0001$ & $0.5351$ & $0.0001$ & $0.1665$ & $0.0002$ & $-4.9610$ & $0.0009$\\
ESPRESSO & $8765.8512$ & $-16.4115$ & $0.0026$ & $6.8855$ & $0.0052$ & $0.3022$ & $0.0003$ & $0.5337$ & $0.0003$ & $0.1553$ & $0.0011$ & $-5.0293$ & $0.0072$\\
ESPRESSO & $8777.8460$ & $-16.4089$ & $0.0013$ & $6.8886$ & $0.0026$ & $0.3058$ & $0.0001$ & $0.5301$ & $0.0001$ & $0.1625$ & $0.0004$ & $-4.9841$ & $0.0024$\\
ESPRESSO & $8787.6432$ & $-16.4099$ & $0.0017$ & $6.8897$ & $0.0033$ & $0.3032$ & $0.0002$ & $0.5293$ & $0.0002$ & $0.1622$ & $0.0007$ & $-4.9859$ & $0.0041$\\
ESPRESSO & $8792.7562$ & $-16.4101$ & $0.0012$ & $6.8816$ & $0.0025$ & $0.3077$ & $0.0001$ & $0.5405$ & $0.0001$ & $0.1611$ & $0.0004$ & $-4.9927$ & $0.0022$\\
ESPRESSO & $8803.7349$ & $-16.4095$ & $0.0009$ & $6.8867$ & $0.0019$ & $0.3078$ & $0.0001$ & $0.5254$ & $0.0001$ & $0.1673$ & $0.0002$ & $-4.9561$ & $0.0013$\\
ESPRESSO & $8806.6179$ & $-16.4045$ & $0.0011$ & $6.8893$ & $0.0021$ & $0.3067$ & $0.0001$ & $0.5280$ & $0.0001$ & $0.1656$ & $0.0003$ & $-4.9662$ & $0.0018$\\
ESPRESSO & $8810.7332$ & $-16.4085$ & $0.0010$ & $6.8846$ & $0.0021$ & $0.3061$ & $0.0001$ & $0.5271$ & $0.0001$ & $0.1666$ & $0.0003$ & $-4.9603$ & $0.0016$\\
ESPRESSO & $8814.6150$ & $-16.4141$ & $0.0009$ & $6.8896$ & $0.0018$ & $0.3065$ & $0.0001$ & $0.5338$ & $0.0001$ & $0.1645$ & $0.0002$ & $-4.9723$ & $0.0013$\\
ESPRESSO & $8817.7561$ & $-16.4072$ & $0.0009$ & $6.8881$ & $0.0018$ & $0.3059$ & $0.0001$ & $0.5321$ & $0.0001$ & $0.1648$ & $0.0002$ & $-4.9707$ & $0.0014$\\
ESPRESSO & $8821.6757$ & $-16.4010$ & $0.0008$ & $6.8859$ & $0.0016$ & $0.3063$ & $0.0001$ & $0.5275$ & $0.0001$ & $0.1654$ & $0.0002$ & $-4.9671$ & $0.0010$\\
ESPRESSO & $8823.7237$ & $-16.4096$ & $0.0008$ & $6.8916$ & $0.0016$ & $0.3077$ & $0.0001$ & $0.5267$ & $0.0001$ & $0.1679$ & $0.0002$ & $-4.9528$ & $0.0010$\\
ESPRESSO & $8838.6116$ & $-16.4016$ & $0.0013$ & $6.8854$ & $0.0026$ & $0.3045$ & $0.0001$ & $0.5273$ & $0.0001$ & $0.1627$ & $0.0004$ & $-4.9830$ & $0.0025$\\
ESPRESSO & $8842.7008$ & $-16.4095$ & $0.0011$ & $6.8889$ & $0.0022$ & $0.3059$ & $0.0001$ & $0.5268$ & $0.0001$ & $0.1643$ & $0.0003$ & $-4.9736$ & $0.0019$\\
ESPRESSO & $8846.6133$ & $-16.4149$ & $0.0012$ & $6.8820$ & $0.0025$ & $0.3057$ & $0.0001$ & $0.5175$ & $0.0001$ & $0.1655$ & $0.0004$ & $-4.9665$ & $0.0022$\\
ESPRESSO & $8858.5827$ & $-16.4110$ & $0.0010$ & $6.8848$ & $0.0021$ & $0.3048$ & $0.0001$ & $0.5300$ & $0.0001$ & $0.1657$ & $0.0003$ & $-4.9656$ & $0.0016$\\
ESPRESSO & $8879.5627$ & $-16.4080$ & $0.0011$ & $6.8905$ & $0.0023$ & $0.3026$ & $0.0001$ & $0.5279$ & $0.0001$ & $0.1616$ & $0.0003$ & $-4.9898$ & $0.0020$\\
ESPRESSO & $8893.5277$ & $-16.4086$ & $0.0011$ & $6.8931$ & $0.0022$ & $0.2972$ & $0.0001$ & $0.4997$ & $0.0001$ & $0.1657$ & $0.0003$ & $-4.9653$ & $0.0019$\\
ESPRESSO & $8911.5011$ & $-16.4092$ & $0.0010$ & $6.8939$ & $0.0020$ & $0.3038$ & $0.0001$ & $0.5274$ & $0.0001$ & $0.1667$ & $0.0003$ & $-4.9594$ & $0.0016$\\
\end{longtable}
\end{landscape}


\bsp	
\label{lastpage}
\end{document}